\documentclass{emulateapj}
\usepackage{natbib}
\usepackage{apjfonts}

\newcommand{\rl}{$R_{\rm BLR}$--$L$}
\newcommand{\msigma}{$M_{\rm BH}-\sigma_{\star}$}
\newcommand{\mbh}{$M_{\rm BH}$}
\newcommand{\hst}{{\it HST}}

\shorttitle{The R--L Relationship: Host-Galaxy Starlight}
\shortauthors{Bentz, et al.}

\received{}
\accepted{}

\begin{document}

\title{The Radius--Luminosity Relationship for Active Galactic
 Nuclei:\\ The Effect of Host-Galaxy Starlight on Luminosity
 Measurements II. The Full Sample of Reverberation-Mapped AGNs}

\author{ Misty~C.~Bentz\altaffilmark{1,2}, 
         Bradley~M.~Peterson\altaffilmark{2,3},
	 Hagai~Netzer\altaffilmark{4},
	 Richard~W.~Pogge\altaffilmark{2,3}, 
	 Marianne~Vestergaard\altaffilmark{5} 
}

\altaffiltext{1}{Present Address:
                 Department of Physics and Astronomy,
		 4129 Frederick Reines Hall,
		 University of California,
		 Irvine, CA 92697;
		 mbentz@uci.edu}

\altaffiltext{2}{Department of Astronomy, 
		The Ohio State University, 
		140 West 18th Avenue, 
		Columbus, OH 43210; 
		peterson, pogge@astronomy.ohio-state.edu}

\altaffiltext{3}{Center for Cosmology and AstroParticle Physics,
                 The Ohio State University,
                 191 West Woodruff Avenue,
                 Columbus, OH 43210}

\altaffiltext{4}{School of Physics and Astronomy and the Wise Observatory, 
                 The Raymond and Beverly Sackler Faculty of Exact Sciences, 
		 Tel-Aviv University, Tel-Aviv 69978, Israel;
		 netzer@wise.tau.ac.il }

\altaffiltext{5}{Department of Physics and Astronomy, 
                 Robinson Hall, Tufts University, 
		 Medford, MA 02155;
		 M.Vestergaard@tufts.edu }

\begin{abstract}

We present high-resolution {\it HST} images of all 35 AGNs with
optical reverberation-mapping results, which we have modeled to create
a nucleus-free image of each AGN host galaxy.  From the nucleus-free
images, we determine the host-galaxy contribution to ground-based
spectroscopic luminosity measurements at $\lambda 5100$\,\AA.  After
correcting the luminosities of the AGNs for the contribution from
starlight, we re-examine the H$\beta$ \rl\ relationship.  Our best fit
for the relationship gives a powerlaw slope of 0.52 with a range of
$0.45 - 0.59$ allowed by the uncertainties.  This is consistent with
our previous findings, and thus still consistent with the naive
assumption that all AGNs are simply luminosity-scaled versions of each
other.  We discuss various consistency checks relating to the galaxy
modeling and starlight contributions, as well as possible systematic
errors in the current set of reverberation measurements from which we
determine the form of the \rl\ relationship.

\end{abstract}


\section{Introduction}

One of the key developments in extragalactic astronomy over the past
decade has been the discovery that supermassive black holes are
present in most, if not all, galaxies having a stellar bulge.
Remarkably, the mass of the black hole is tightly correlated with the
stellar velocity dispersion of the host galaxy bulge
(\citealt{ferrarese00,gebhardt00}), pointing to a close link between
the growth and evolution of galaxy stellar populations and the growth
of nuclear black holes.  Determining the masses of black holes in
active galaxies is a crucial step toward understanding this
connection, and provides fundamental insight into the physics of
accretion and emission processes in the black hole environment.

Unfortunately, most active galactic nuclei (AGNs) are too distant for
black hole masses to be measured using spatially resolved stellar or
gas dynamics.  The technique that has been most successful for the
measurement of the black hole mass (\mbh) in AGNs is reverberation
mapping (\citealt{blandford82,peterson93}).  With this technique, the
AGN continuum (typically measured at 5100\,\AA) and broad emission
lines (most notably, H$\beta$) are monitored over an extended time
period.  Since the emission-line regions are photoionized by the
central source, changes in the AGN continuum strength are followed by
changes in the emission-line fluxes, with a time lag that depends
on the light-travel time across the broad-line region (BLR).  This
time lag can be measured by cross-correlation of the continuum and
emission-line light curves, and gives the radius of the BLR.
Combining the BLR radius with the broad emission-line velocity width
then gives the virial mass enclosed within the BLR, which is dominated
by the black hole (e.g., \citealt{peterson98a,kaspi00}).  The validity
of reverberation masses has been upheld by the detection of virial
behavior in the broad line region in a subset of objects (e.g.,
\citealt{peterson99,peterson00a,onken02,kollatschny03}), as well as
the consistency of reverberation masses with other dynamical mass
methods, such as stellar dynamics (\citealt{davies06,onken07}) and gas
dynamics \citep{hicks08}.  Due to the long-term nature of
reverberation-mapping projects, these measurements have only been
carried out for a relatively small sample of AGNs in the past: about
36 Seyferts 1s and low-luminosity quasars.

The BLR radius--luminosity correlation ($R_{\rm BLR} \propto
L^{\alpha}$) derived from this reverberation sample is the basis for
\emph{all} secondary techniques used to estimate black hole masses in
distant AGNs (e.g., \citealt{laor98,wandel99,mclure02,vestergaard06})
and is an essential tool used to search for cosmological evolution of
the \msigma\ relationship (e.g., \citealt{peng06,woo08}).  The power
of the \rl\ relationship comes from the simplicity of using it to
quickly estimate \mbh\ for large samples of objects, even at high
redshift, using two simple measurements from a single spectrum of each
object.

\citet{peterson04} compiled and consistently reanalyzed the database
of available reverberation-mapping data for 35 AGNs, with the goal of
improving the measurements of the size of the BLR and thereby
improving their mass measurements.  Subsequently, \citet{kaspi05}
reanalyzed the \rl\ relationship and found a power-law slope of
$\alpha = 0.665 \pm 0.069$ using the optical continuum and broad
H$\beta$ line.  However, many of the AGNs in the sample reside in host
galaxies that are comparable in luminosity to the AGN itself.  Even
worse, with the typically large apertures employed in
reverberation-mapping campaigns (i.e. 5$\farcs$0 $\times$ 7$\farcs$5),
the host-galaxy starlight contribution to the spectroscopic luminosity
of any given source is substantial.  Failure to account for the
enhancement of the luminosity by starlight results in an artificially
steep slope for the \rl\ relationship, as a larger percentage of the
luminosity in faint AGNs is contributed by the host galaxy.

A preliminary study by \citet{bentz06a} presented two-dimensional fits
to high-resolution \hst\ images of 14 reverberation-mapped AGNs, from
which the host-galaxy contribution was determined.  The luminosities
of the 14 sources were corrected and the \rl\ relationship was
re-examined, resulting in a measured power-law slope of $\alpha
\approx 0.5$, consistent with the naive prediction that all AGNs are
simply luminosity-scaled versions of each other.  In this work, we
present high-resolution \hst\ images of the rest of the H$\beta$
reverberation-mapped AGNs, bringing the total to 35. We improve upon
our previous two-dimensional fits and reanalyze the \rl\ relationship
after correcting the luminosity of every AGN for the contribution from
starlight.  We show that these new results are consistent with those
from our preliminary study, and discuss the new measurements in light
of known systematic errors that may affect the slope of the \rl\
relationship.  Throughout this work, we will assume a standard flat
$\Lambda$CDM cosmology with $\Omega_{\rm B} = 0.04$, $\Omega_{\rm DM}
= 0.26$, $\Omega_{\Lambda} = 0.70$, and $H_0 =
70$\,km\,s$^{-1}$\,Mpc$^{-1}$.

\section{Observations and Data Reduction}

\subsection{Hubble Space Telescope}

Between 2003 August 22 and 2007 January 17, we observed 30 AGNs from
the re\-ver\-ber\-a\-tion-mapped sample of \citet{peterson04} with the
{\it HST} Advanced Camera for Surveys (ACS).  Following the failure of
ACS on 2007 January 27, the remaining 5 AGNs in our sample were
observed with Wide Field Planetary Camera 2 (WFPC2).  The targets are
listed in Table~1 and details of the observations are listed in
Table~2.

For the ACS observations, each object was imaged with the High
Resolution Channel (HRC) through the F550M filter ($\lambda_{c} =
5580$\,\AA\ and $\Delta \lambda = 547$\,\AA), thereby probing the
continuum while avoiding strong emission lines.  The observations
consisted of at least three exposures for each object, with exposure
times of 120\,s, 300\,s, and 600\,s.  This method of graduating the
exposure times was employed to avoid saturation of the nucleus but
still obtain a reasonable signal-to-noise ratio ($S/N$) for the wings
of the point-spread function (PSF) and the host galaxy.  Each
individual exposure was split into two equal sub-exposures to
facilitate the rejection of cosmic rays.  The WFPC2 observations were
centered on the PC chip and were taken through the F547M filter, the
closest analog to the ACS F550M filter. The exposure times were again
graduated, however, we used steps of 5\,s, 20\,s, 60\,s, 160\,s and
300\,s due to the smaller time-to-saturation afforded by the pixels in
the PC chip.

The data quality frames provided by the {\it HST} pipeline were
consulted to identify the individual saturated pixels associated with
the nucleus in each exposure frame.  These saturated pixels were clipped
from the image and replaced by the same pixels from a non-saturated
exposure after scaling them by the relative exposure times.  All of the
frames for each object were then summed to give one frame with an
effective exposure time as listed in Table~2.

Cosmic rays were identified in the summed images with the Laplacian
cosmic ray identification package L.A.Cosmic \citep{vandokkum01}.
Pixels in the PSF area of each image that were identified by
L.A.Cosmic were excluded from the list of affected pixels prior to
cleaning with XVista.\footnote{XVISTA was originally developed as Lick
Observatory Vista and is now maintained in the public domain by former
Lick graduate students as a service to the community.  It is currently
maintained by Jon Holtzman at New Mexico State University, and is
available at http://ganymede.nmsu.edu/holtz/xvista.}  Each remaining
affected pixel was replaced with the median value for the eight pixels
immediately surrounding it.

Finally, the summed, cleaned ACS images were corrected for the
distortions of the camera using the PyRAF routine {\it pydrizzle} in
the STSDAS\footnote{STSDAS and PyRAF are products of the Space
Telescope Science Institute, which is operated by AURA for NASA.}
package for IRAF.  The final stacked, cleaned images for all 35 AGNs
are shown in Figure~1, overlaid with the spectroscopic aperture
geometries from their ground-based monitoring campaigns.

\subsection{MDM Observatory}

Images of the reverberation-mapped sample of galaxies were also taken
with the 1.3-m McGraw-Hill Telescope at MDM Observatory. Templeton, a
$1024 \times 1024$ pixel CCD, was employed for the observations,
giving a field-of-view (FOV) of \mbox{8\farcm 53 $\times$ 8\farcm 53}
and a pixel scale of 0.50\arcsec~pixel$^{-1}$.  Each galaxy was imaged
through Harris $B$, $V$, and $R$ filters.  We focus here on the
observations of the seven NGC objects that were visible from the
location of MDM Observatory\footnote{NGC~3783 is located at a
declination of $-38\degr$ and was therefore not observed.}.  A log of
the MDM observations for those objects is presented in Table~3.  The
data were reduced and combined with IRAF\footnote{IRAF is distributed
by the National Optical Astronomy Observatory, which is operated by
the Association of Universities for Research in Astronomy (AURA) under
cooperative agreement with the National Science Foundation.} following
standard procedures.  The final $V$-band images are shown in Figure~2.

\section{Galaxy Decompositions}

The images of each of the objects were modeled with typical galaxy
parameters in order to determine and accurately subtract the
contribution from the central point source.  The models were
constructed using the two-dimensional image decomposition program
Galfit \citep{peng02}, which fits analytic functions for the
components of the galaxy, plus an additional point source for the
nucleus, convolved with a user-supplied model point-spread function
(PSF).

For each of the objects in this study, the final cleaned, stacked
image was fit with a central PSF and a constant sky contribution, as
well as host-galaxy components that were modeled using variations of
the \citet{sersic68} profile,
\begin{equation}
\Sigma (r) = \Sigma_e \exp^{-\kappa [(r/r_e)^{1/n}-1]} 
\end{equation}
where $r_e$ is the effective radius of the component, $\Sigma_e$ is
the surface brightness at $r_e$, $n$ is the power-law index, and
$\kappa$ is coupled to $n$ such that half of the total flux is within
$r_e$.  Two special cases of the S\'{e}rsic function are the
exponential profile ($n=1$), often used in modeling galactic disks,
and the \citet{devaucouleurs48} profile ($n=4$), historically used for
modeling galactic bulges.  We modeled disk components using the
exponential profile.  However, we improve upon the results presented
by \citet{bentz06a} in that we employed the more general S\'{e}rsic
function for modeling bulges and we allow for additional parameters to
describe other surface brightness components (such as a bar or inner
bulge component).  These modifications were partially prompted by the
large body of observations that find disk galaxies (of which our
sample is mostly comprised) are more accurately described with bulge
profiles that have $n<4$
(e.g. \citealt{kormendy78,shaw89,andredakis94,peng02}).

There is a paucity of archival stellar images with the HRC through the
F550M filter, and so simulated PSFs were created using the TinyTim
package \citep{krist93} which models the optics of \hst\ plus the
specifics of the camera and filter system.  We tested our fits using a
white dwarf image from the \hst\ archive as the PSF model (GO 10752,
PI Lallo).  Unfortunately, the image did not have the extremely high
dynamic range necessary for a good fit when compared to these bright
AGNs.

Many of the nearest galaxies in this sample fill the FOV of the ACS
HRC, and so the sky contribution could not simply be measured from the
edges of the images.  Rather, the sky level was iteratively
determined, starting with an estimate of the sky brightness from the
ACS Instrument Handbook as the initial input.  The sky value was held
fixed while additional parameters were fit to the galaxy.  If the
estimated sky value was too low, the S\'{e}rsic index for the bulge
would run up to the maximum allowed value, punching a ``hole'' in the
nucleus of the image.  If the estimated sky value was too high, the
S\'{e}rsic index could run down to zero, causing Galfit to crash.  A
sky value intermediate to these two situations was chosen such that
the residuals and $\chi^2$ values were minimized.  Once a preliminary
fit was achieved, the sky value was checked and adjusted if necessary,
after which the galaxy parameters were re-fit.  This is an improvement
over our previous work, where the sky contribution was assumed to be
negligible compared to the bright host galaxies of the 14 AGNs in that
sample, especially as our expanded sample includes several AGNs that
are significantly brighter than their host galaxies.

All of the images required at least one galaxy component in addition
to the sky and central PSF.  Most required two, and a few required
three or more to fit an additional component, such as a bar.  Several
of the bright AGNs required a small ($ < 2$~pixel effective radius)
component to help account for PSF mismatch between the AGN and the
model PSF (for a full discussion of PSF variations and mismatches in
{\em HST} imaging, see \citealt{krist03,kim08}).  Extraneous objects
in the field were also fit, such as intervening stars or galaxies, to
ensure that the proper distribution of light was attributed to the AGN
host galaxy.  Compared to the simple galaxy decompositions for the 14
objects in our previous work, the surface brightness residuals and
nominal $\chi^2$ fitting values have been significantly reduced,
showing that the fits presented here more accurately model the
underlying host galaxy surface brightness distributions.  For example,
the average magnitude of the residuals for 3C\,120 was reduced by a
factor of $\sim 4$ and the $\chi^2$ value was reduced by a factor of
$\sim 5$.\footnote{The actual value of $\chi^2$ is not particularly
meaningful when related to galaxy fitting.  The relative $\chi^2$
values for different fits to the same image are more relevant, with
smaller values denoting ``better'' fits.  Details of the calculation
and interpretation of $\chi^2$ values can be found in the FAQ section
of Chien Peng's Galfit home page,
http://users.ociw.edu/peng/work/galfit/galfit.html} In all cases, the
fits have been encouraged to attribute more flux to the sky background
and PSF components, resulting in conservative flux values for the
host-galaxy components.  The quoted brightness for each of the host
galaxy components may be somewhat underestimated as a result.

For the seven NGC objects with MDM images, the $V$-band MDM images
were each fit with an exponential disk and a S\'{e}rsic bulge.  The
effective radius of the disk component was then translated to the
pixel scale of the HRC camera and held fixed during the fits to the
{\it HST} image.  As the seeing in the MDM images was typically on the
order of $\sim 2\arcsec$, the PSF and the bulge of the galaxy were
blurred together.  Thus, the bulge fits from the ground-based images
were held to be unreliable and were instead determined from the \hst\
images. 

Table~4 presents the details of the fits to the \hst\ images.  The
input image, Galfit model, residuals, and one-dimensional surface
brightness cut for each of the 35 host galaxies are presented in
Figure~3.  Global parameters including the total galaxy luminosity and
the ratio of the bulge luminosity to the total luminosity ($B/T$) were
determined from the fits and are listed in Table~5.  Also listed in
Table~5 are the morphological classifications of the galaxies, several
of which were listed in the NASA/IPAC Extragalactic Database (NED).
For those objects without morphological classifications (objects
marked with a flag in Table~5), the parameters fit to the galaxy
images were used to determine the appropriate classification based on
the \citet{devaucouleurs59} classification scheme.  The index $k$ for
the subtype of the elliptical galaxies in the sample was calculated as
$k = 10(1 - b/a)$ and rounded to the nearest whole number.  Spiral
galaxies were classified based on their $B/T$ compared to the mean of
the distributions of $B/T$ as a function of morphological type in
Figure~6 of \citet{kent85}.

As discussed by \citet{peng02}, degeneracy between galaxy components
and between parameters within components is typically an issue when
fitting analytic models to galaxy images.  This is certainly the case
with the sample of objects presented here.  The bright AGNs in the
galaxy centers are degenerate with the concentration of the bulge (the
S\'{e}rsic index $n$), especially when the bulge component is rather
compact.  As discussed above, many of these images have the added
complication of having a somewhat uncertain sky contribution which
also affects $n$ as well as the scale length of the disk.
Interpreting the specific details of the galaxy fits presented here is
therefore difficult.  The relatively low $n$ values for the galaxy
bulges in this sample certainly seem to agree with the works
referenced above that find $n<4$ for most spiral galaxies.  However,
the actual $n$ values may be somewhat underestimated as a result of
our conservative fits that attribute more flux to the AGN and the sky
background.  A quick glance through the galaxy fits in Table~4 might
lead one to speculate on the prevalence of pseudobulges versus
classical bulges in this sample.  Per the discussion by
\citet{kormendy04}, however, the lack of dynamical info for the host
galaxies in this sample as well as the uncertainty in interpreting the
galaxy fitting parameters complicates any conclusions that might be
drawn about the origins of the bulges in these galaxies.

Several of the fits listed in Table~4 include components noted as
being an ``inner bulge'' or ``bar''.  The classifications for these
extra components are simply morphological, with a ``bar'' typically
having a more elongated shape than an ``inner bulge''.  Bars
themselves tend to have an inner bulbous component, so the ``inner
bulges'' listed here may actually be related to bars themselves
\citep{peng02}.  We do not have kinematic information to determine
which of these additional components may be physically distinct from
the bulge or disk of the host galaxy.  The fact that additional
components are required to achieve a better fit is likely in many
cases to simply be the result of attempting to fit analytic functions
to resolved galaxies with strong substructure.  In particular, areas
with dust obscuration and spiral arm structure are difficult to
properly fit with analytic functions.  As our main goal is to
determine the PSF contribution as accurately as possible and to create
``nucleus-free'' images of these galaxies, we do not discuss here the
origin or meaning of these additional components in the galaxy fits.

\section{Flux Measurements}

Once the fit to each galaxy was finalized, the sky and PSF components
were subtracted, leaving a nucleus-free image of each host galaxy.
Each image was overlaid with the typical aperture used in the
ground-based monitoring program(s), at the typical orientation and
centered on the position of the AGN (see Table~6).  The counts within
the aperture were summed and converted to $f_{\lambda}$ flux density
units (ergs~s$^{-1}$~cm$^{-2}$~\AA$^{-1}$) using the \hst\ keyword
PHOTFLAM and the effective exposure time for each object. 

Color corrections to the observed galaxy flux densities were
calculated using a model bulge spectrum \citep{kinney96} to account
for the difference between the effective wavelength of the \hst\
filter and restframe 5100\,\AA\ for each object.  The model bulge
spectrum was redshifted to the distance of each AGN host galaxy and
reddened by the appropriate Galactic extinction.  The redshifted,
reddened models were convolved with the \hst\ filter response using
the {\it synphot} package in the IRAF/STSDAS library and subsequently
scaled to the appropriate flux level as measured from the \hst\
images.  Finally, the observed flux at rest-frame 5100\,\AA\ was
measured from the model.  The flux density measured directly from the
nucleus-free \hst\ images and the color-corrected flux and luminosity
at 5100\,\AA, are listed in Table~7.  We note that the color
correction method outlined above is different from that employed
previously, and again results in more conservative host galaxy flux
values.

The final step requires subtraction of the galaxy flux from the mean
flux measured during a monitoring campaign, and leaves only the flux
contribution from the AGN itself.  The corresponding AGN luminosities
were calculated and corrected for Galactic absorption using the
\citet{schlegel98} $A_B$ values listed in NED and the extinction curve
of \citet{cardelli89}, adjusted to $A_V/E(B-V) = 3.1$.  The
starlight-corrected AGN fluxes and luminosities are listed in Table~8
with their corresponding H$\beta$ lags.  The numbers in bold font are
the weighted averages of multiple measurements for a particular
object.

Ongoing work with this sample of objects has resulted in some updates
to the reverberation database presented by \citet{peterson04} and fit
by \citet{kaspi05}.  In addition, a few data sets require cautionary
or explanatory notes.  Specifically, we note the following:
\begin{itemize}

\item NGC\,3516: The light curves presented by \citet{wanders93} were
not measured from data with an absolute flux calibration.  Therefore,
while we have included this object, it should be regarded with
caution as the spectroscopic luminosity is not well-determined.

\item NGC\,4151: The aperture used in the monitoring campaign
described by \citet{maoz91} is 20\arcsec $\times$ 28\arcsec, which is
larger than the field of view of the HRC camera.  In addition, the
\citet{kaspi96} data set has a rather unconstrained time lag that is
consistent with zero once the monotonic increase in the continuum and
line flux is removed (see \citealt{metzroth06} for a full discussion).
Both of these data sets for NGC\,4151 have been superseded by the
results reported by \citet{bentz06b}, which we include here in their
stead.

\item PG\,1211+143: \citet{peterson04} note that all of the lag
measurements for this object are rather unconstrained.  For
consistency with past analyses of the \rl\ relationship, we have
included the H$\beta$ lag for PG1211 here, but it should be regarded
with caution.

\item NGC\,4593: The campaign originally described by
\citet{dietrich94} and reanalyzed by \citet{onken03} gives an H$\beta$
lag that is consistent with zero.  The results for NGC\,4593 have been
superseded by those reported by \citet{denney06}, which we include
here.

\item IC\,4329A: According to \citet{peterson04}, the light
curves for this object are very noisy and of poor quality, resulting
in an H$\beta$ lag measurement that is highly suspect as well as being
consistent with zero.  In addition, IC\,4329A is an edge-on galaxy
with a prominent dust disk along the line of sight to the AGN,
resulting in a substantial amount of internal reddening.  The presence
of the strong dust lane across the galaxy in the {\it HST} images also
presented problems for the host galaxy fitting.  Because this object
has both an unreliable lag measurement and an unreliable luminosity
measurement, we have excluded it from the analysis of the \rl\
relationship.

\item Mrk\,279: The monitoring campaign described by \citet{maoz90}
used an aperture of 20\arcsec $\times$ 28\arcsec, which exceeds the
HRC field of view.  We include here only the results for Mrk\,279
reported by \citet{santoslleo01}.

\item NGC\,5548: The aperture of 20\arcsec $\times$ 28\arcsec\ used in
the campaign described by \citet{netzer90} is larger than the field of
view of the HRC camera.  We include here the 14 individual
measurements reported by \citet{peterson02} and \citet{bentz07}.

\item PG\,2130+099: For some time now, the H$\beta$ lag measured for
this object has been suspected of being artificially long.  This
suspicion has been confirmed by \citet{grier08}, who have recently
reanalyzed the data set originally presented by \citet{kaspi96} and
find that the lag measured from these data is erroneous due to
undersampling of the light curves and long-term secular changes in the
equivalent width of the broad H$\beta$ emission line.  In addition,
\citeauthor{grier08} present a new monitoring campaign for this object
that was undertaken at MDM Observatory in the fall of 2006, and report
a new H$\beta$ lag measurement. We include only the
\citeauthor{grier08} result in our analysis here.

\end{itemize}

\section{Consistency Checks}

The difficulty of fitting surface brightness profiles with multiple
analytic functions is widely appreciated. While the GALFIT program we
have used works remarkably well and represents the state of the art,
there are still some ambiguities that cannot always be easily
resolved: at a sufficiently large distance, for example, it is simply
impossible, even with the high angular resolution of the ACS HRC, to
separate a point source AGN and a luminous bulge with a high
S\'{e}rsic index. In addition, several parameters that are used in the
surface brightness fits can be very degenerate, leading to
difficulties in disentangling the results from many different
parameter combinations when no other information is available.  All
this may result in systematic uncertainties that are larger than the
formal errors (of order 10\%) that are listed in Table~7.  There are,
however, a number of consistency checks that provide robust upper
limits on the host galaxy contributions and can help to remove some of
this degeneracy.  Indeed, our decision to revisit the \citet{bentz06a}
fits was motivated at least in part by such considerations.

The simplest consistency check is that the AGN flux must always remain
non-negative: the host galaxy flux at any wavelength must be less than
the observed brightness of the combined AGN and host galaxy when the
AGN is in its faintest observed state.  We have compared the measured
host galaxy fluxes to the individual monitoring light curves for each
of the objects in this sample to ensure that this consistency check is
always satisfied.  A second consistency check is provided by spectral
decomposition, i.e., by fitting multiple components of a spectrum
rather than an image. Our experience, however, is that spectral
decomposition is certainly no less ambiguous than image decomposition,
but when carefully done with very good data, shows reasonable
consistency (e.g., the modeling of the optical spectrum of NGC\,5548
by \citealt{denney08} yields a host galaxy flux in agreement with the
value we find here). A third check, which applies at least in the case
of the higher luminosity AGNs, is that the total luminosity of the
host galaxy, based on the model fit, must not exceed the luminosity of
known bright normal galaxies.  Our galaxy fits seem to be consistent
with this constraint as well.  The brightest host galaxy in the sample
is PG\,1226+023, also known as 3C\,273, which has $M_V \approx -23.8$,
and the second brightest with $M_V \approx -23.2$ is PG\,1700+518, the
most distant AGN in the sample at $z=0.292$.  Compared to the
\citet{trentham05} field galaxy luminosity function, the derived host
galaxy brightnesses for the AGNs in this study seem to be fairly
typical.

A more subtle consistency check on the host-galaxy contribution to the
ground-based monitoring luminosity comes from photoionization
considerations, namely that the ratio of emission-line to AGN
continuum flux (i.e., the emission-line equivalent width) ought not to
vary wildly in time.  For an optically thick medium, the equivalent
width of an emission line is expected to stay relatively constant.
Some variations are possible, either due to changes in the continuum
shape or if there is an optically thin BLR component.  Significant
variations in the equivalent width can follow a large-amplitude change
in the incident continuum flux, however, the variations should
stabilize and return to a relatively constant value shortly
thereafter.  In addition, secular trends in the equivalent width can
occur over dynamical timescales, which are much longer than
reverberation timescales.  If too much flux is attributed to the host
galaxy of an object, the AGN continuum will be underestimated and the
fractional variations of the H$\beta$ emission-line equivalent width
will become consistently large, in contrast to the variations that can
be expected from the discussion above.  An important point to note is
that the relevant equivalent width measure is the ratio of
emission-line flux at some time, $t$, relative to the continuum flux
at some earlier time, $t - \tau$, where $\tau$ is the emission-line
time delay from reverberation measurements (e.g.,
\citealt{pogge92,gilbert03}).

To illustrate this check, we consider the specific case of 3C\,120,
which we selected because the starlight contribution changed the most
from our previous work, decreasing by a factor of almost 3. In
Figure~4 we show the equivalent width of the H$\beta$ emission line
(in the observed frame) as a function of time, based on the light
curve from \citet{peterson98a}. Each measurement of the emission-line
flux, $F_{{\rm H}\beta}(t)$, was divided by an interpolated value of
the continuum flux density, $F_{\lambda}(t-\tau)$, with an
observed-frame emission-line lag of $\tau = 39.4$\,days.  We note in
passing that the first few emission-line measurements are disregarded
since there is no corresponding continuum information.  The open
circles in Figure~4 refer to the original data, without any correction
for starlight.  The triangles are the new equivalent width
measurements for the same emission line fluxes, but in this case the
continuum measurements have been adjusted by the host-galaxy value
from \citet{bentz06a}.  The large equivalent widths and consistently
rapid variations strongly suggest that this value of the host-galaxy
correction is too large, as the fractional variations of the remaining
AGN continuum must be enormous. The filled circles show the equivalent
widths based on the revised value of the host-galaxy contribution
given here in column (3) of Table~7. In this case, the values of the
equivalent widths are much more reasonable (i.e., typical of quasars,
where host-galaxy contamination is negligible) and the variations are
much more moderate, indicating that the host-galaxy measurement
provided here is more accurate than the measurement based on the more
simplistic host-galaxy models utilized by \citeauthor{bentz06a}

For all 35 objects in the sample, the galaxy fits in \S3 were carried
out independently from the consistency checks discussed above.  The
more conservative approach we have applied to the galaxy fits here
results in good agreement with the expectations from all of these
consistency checks for the 35 objects in the sample.  Of the 14
objects included in our preliminary study, three objects have
starlight contributions through their monitoring apertures that
changed by more than 35\% as measured directly in counts from the
nucleus-free images, one of them being 3C\,120, as discussed above.
In all cases, the decreased residuals and $\chi^2$ values from the new
fits, combined with the results of the consistency checks, lead us to
believe that the galaxy fits in this work more accurately describe the
underlying host-galaxy surface brightness distributions than the
simplistic fits determined by \citet{bentz06a}.

\section{The Radius--Luminosity Relationship}

We have calculated several fits to the \rl\ relationship for the full
sample of starlight-corrected AGNs.  For this analysis, the time lags
have been restricted to the H$\beta$ lag only.  This is different from
the method employed by \citet{kaspi05}, where the \rl\ fits were also
calculated for the Balmer-line average, which sometimes included
H$\alpha$ and/or H$\gamma$.  There is one exception to this, however,
in that PG\,0804+761 does not have a reliable H$\beta$ measurement,
but it does have reliable measurements for H$\alpha$ and H$\gamma$.
We use the lag measurement for H$\alpha$ here, as it is the more
reliable of the two measurements available.  

In the same way as \citeauthor{kaspi05}, we have made the distinction
of treating each separate H$\beta$ lag measurement of an object
individually, as well as taking the mean of multiple measurements
weighted by the average of the positive and negative errors.  We
tested the differences between weighting measurements by the average
of their errors, by taking only the positive errors, and by taking the
errors toward the fit in the manner of \citet{kaspi05}.  We find the
differences in these weighting methods to be at the 2$\%$ level, and
therefore negligible.

While the two above methods for sampling the reverberation-mapping
database are relatively straightforward to carry out, various issues
arise when determing the \rl\ relationship from these datasets.  In
the first case, where every measurement is given equal weight, those
objects that have many measurements (such as NGC\,5548) will have more
weight in the determination of the slope than objects with single
measurements.  In the second case, where multiple measurements are
averaged together, each object has the same weight but information is
being lost in the average because we do not, in fact, expect the
luminosity and lag to be the same in multiple campaigns during
different years.  It is unclear as to the correct way to combine
multiple measurements in this case.

A somewhat more laborious method of sampling the reverberation-mapping
database is to randomly select one pair of radius and luminosity
measurements for every object, from which selection the relationship
is fit, and to build up a large number ($N=1000$) of individual
realizations through Monte Carlo techniques.  This method gives equal
weight to every object and circumvents the problem of how to combine
multiple measurements for any particular object.  In addition to the
two simpler methods outlined above, we employ this method in the
determination of the \rl\ relationship, but we consider this method to
be superior to the others.

Three different fitting routines were used to calculate the
relationship between the size of the BLR and the optical luminosity:

\begin{description}
\item[1.] FITEXY \citep{press92}, which estimates the parameters of a
        straight-line fit through the data including errors in both
        coordinates.  FITEXY numerically solves for the minimum
        orthogonal $\chi^2$ using an interative root-finding
        algorithm.  We include intrinsic scatter similar to
        \citet{kaspi05} after the prescription of \citet{tremaine02}.
        Namely, the fractional scatter listed in Table~9 is the
        fraction of the measurement value of $R_{\rm BLR}$ (not the
        error value) that is added in quadrature to the error value so
        as to obtain a reduced $\chi^2$ of 1.0.
\item[2.] BCES \citep{akritas96}, which attempts to account for the
        effects of errors on both coordinates in the fit using
        bivariate correlated errors, including a component of
        intrinsic scatter.  We adopt the bootstrap of the bisector
        value following \citet{kaspi05}, with $N=1000$
        iterations.\footnote{We note that the errors on $L$ listed in
        the table are formal measurement errors that, in most cases,
        are far smaller than the actual observed variations even in
        one campaign. Therefore, we also tested BCES correlations that
        assume an uncertainty of 0.1-0.15\,dex on $L$. The results are
        similar to the ones obtained by our realization method and
        hence not listed separately.}
\item[3.] GaussFit \citep{mcarthur94}, which implements generalized
	least-squares using robust Householder Orthogonal
	Transformations \citep{jefferys80,jefferys81} to solve the
	non-linear equations of condition for the problem of errors in
	both coordinates.  No attempt is made to account for intrinsic
	scatter.
\end{description}

\noindent Table~9 lists the fit parameters determined for a \rl\ relationship
of the following form:
\begin{equation}
{\rm log}\,(R_{\rm BLR}) = K + \alpha\ {\rm log}\,(\lambda L_{\lambda}\,
(5100 {\rm{\mbox \AA}}))
\end{equation}
where $\alpha$ is the slope of the power-law relationship between
$R_{\rm BLR}$ and $\lambda L$\,(5100\,\AA), and $K$ is the zero point.
The calculated power-law slopes to the \rl\ relationship range from
$0.499 \pm 0.042$ to $0.554 \pm 0.050$, depending on the particular
alorithm used for fitting the relationship and how the objects with
multiple measurements are treated.  

We prefer the Monte Carlo random sampling method outlined above as the
proper way to treat objects with multiple individual measurements.
And for its manner of dealing with intrinsic scatter within the data
set, we prefer the BCES bootstrap method.  The combination of these
methods gives $K = -21.3^{+2.9}_{-2.8}$ and $\alpha =
0.519^{+0.063}_{-0.066}$, our best estimate for the form of the \rl\
relationship at this time.

Figure~5 shows the \rl\ relationship after correcting the full sample
of reverberation-mapped AGNs for the contribution from host-galaxy
starlight.  The top panel of Figure~5 shows each individual data point
from the monitoring campaigns included here, and the bottom panel
shows a single data point for each individual AGN determined by the
weighted average of multiple measurements.  The solid lines show the
best fit to the \rl\ relationship described above.

\section{Discussion}

Throughout this work, we have improved upon our original methods by
using more accurate and more conservative profiles to model the host
galaxy components, as well as a more conservative color correction
method, all resulting in conservative measurements of the host galaxy
starlight as measured for every object contributing to the \rl\
relationship.  Even so, the best fit to the relationship has not
changed significantly from that presented by \citeauthor{bentz06a}:
$0.519^{+0.063}_{-0.066}$ compared to the previous value of $0.518 \pm
0.039$.  It would appear that given reasonably high-resolution,
unsaturated images of AGN host galaxies and reasonable fits to the
host galaxy surface brightness profiles, the AGN luminosity can be
corrected for the contamination from starlight fairly accurately.

AGN BLRs can be modeled assuming the central radiation field is the
only source of heating and ionization of the gas. The simplest, most
naive assumption is that all BLRs are made of identical clouds, with
the same density, column density, composition and ionization
parameter.  In addition, the radiation from all AGNs can be assumed to
have the same spectral energy distribution (SED).  This results in the
prediction that $R_{\rm BLR} \propto L^{0.5}$.  Any changes in the BLR
gas distribution or the AGN SED, especially those associated with
source luminosity, will result in a different slope.  In fact, AGNs
have been observed to have different SEDs as a function of luminosity
(e.g., \citealt{mushotzky89,zheng93}).  It is therefore interesting
that our present work based on the results from several
reverberation-mapping campaigns produces a slope for the \rl\
relationship that is consistent with the naively expected slope of
0.5, and that, to a first approximation, brighter AGNs are simply
``scaled up'' versions of fainter AGNs.

Removing the host-galaxy starlight component reduces the
scatter\footnote{Throughout this manuscript, ``scatter'' refers to the
$1-\sigma$ deviation from the best fit \rl\ relationship.} in the \rl\
relationship, from $39-44$\% as found by the best-fit FITEXY results
from \citet{kaspi05} to $34-40$\% as found here.  It also flattens the
slope of the relationship considerably.  This has the overall effect
of biasing samples that used previous versions of the \rl\
relationship to estimate black hole masses.  The host-galaxy starlight
is often removed through spectroscopic decomposition of single-epoch
spectra before the size of the BLR radius is estimated from the
continuum luminosity.  However, it is crucial to use a \rl\
relationship in which the objects providing the calibration are {\it
also} corrected for host-galaxy starlight, which we have provided
here.  Compared with the best-fit relationship of \citet{kaspi05}, we
find that the calibration of the \rl\ relationship presented here
results in black hole masses that are $\sim 30$\% smaller at $L =
10^{46}$, $\sim 50$\% larger at $L = 10^{44}$, and a factor of $\sim
3$ larger at $L = 10^{42}$.

At the low-luminosity end ($L \la 10^{43}$), there may still be some
uncertainty as to the behavior of the \rl\ relationship.  In the
current sample of objects with reveberation-mapping results, there are
somewhat fewer objects at lower-luminosity and they have larger
uncertainties than the other, higher-luminosity objects in the sample.
It should be kept in mind that the lower-luminosity objects were, in
general, the first targets of ground-based monitoring campaigns due to
their relatively low redshifts and high apparent brightnesses.  The
larger uncertainties in their measurements is partially due to the
less-rigorous control over observational factors in those early
monitoring campaigns (such as observing cadence, spectral resolution,
detector efficiency, etc.) simply because there was a lack of
experience in this field at that time.  The problems of small sample
size and relatively larger uncertainties for lower luminosity objects
will soon be mitigated by two independent reverberation mapping
campaigns which have been recently carried out at MDM Observatory and
at Lick Observatory and targeted the low-luminosity end of the
relationship.  Preliminary results from the MDM campaign promise to
replace several measurements, as was done in the case of NGC\,4593
\citep{denney06} and NGC\,4151 \citep{bentz06b} in the 2005 MDM
campaign. And preliminary results from the Lick campaign promise to
add several new objects to the low-luminosity end of the \rl\
relationship (e.g., \citealt{bentz08}).

Internal reddening is known to be a problem in some of the very
nearest, and lower-luminosity, objects in the current reverberation
sample.  For example, NGC\,3227 is known to have substantial internal
reddening compared to most of the other objects in the current sample
and should therefore require one of the largest reddening corrections.
\citet{crenshaw01} determined a reddening curve for NGC\,3227 and find
that at 5100\,\AA, $A_{\lambda} / E(B-V) = 3.6$ and $E(B-V) = 0.18$.
This implies an extinction at 5100\,\AA\ of 0.65\,mag, which, if
corrected for, would increase the luminosity here by a factor of 1.8,
or 0.26\,dex, moving the location of NGC\,3227 in the bottom panel of
Figure~5 from slightly left of the best fit \rl\ relationship to right
on top of it.  As we do not have similar corrections for the other
objects in this sample, we do not apply the reddening correction for
NGC\,3227 in our determination of the \rl\ relationship.  However,
based on the magnitude of the correction determined for NGC\,3227,
there does not seem to be any reason to expect that correcting all the
sources for internal reddening will have much effect on the slope of
the \rl\ relationship.

An obvious issue that may also be addressed with the galaxy fits that
we have presented in this work is the relationship between black hole
mass and host galaxy bulge luminosity (or mass;
\citealt{kormendy95,magorrian98}).  We discuss this relationship for
the AGNs in our sample in a related manuscript \citep{bentz09}.

\section{SUMMARY}

We have presented high-resolution {\it HST} images of the 35 AGNs with
optical reverberation-mapping results.  The host galaxy of each object
was fit with typical galaxy components, and a nucleus-free image of
each AGN host galaxy was created.  From these nucleus-free images, we
measured the starlight contribution to the ground-based spectroscopic
luminosity measured at rest-frame 5100\,\AA.  We then removed the
starlight contamination from the AGN luminosities and re-examined the
\rl\ relationship.  We find a best fit slope of $\alpha =
0.519^{+0.063}_{-0.066}$, consistent with the results from our
preliminary study, and still suggesting that all AGNs are simply
luminosity-scaled versions of each other.  We discuss several
consistency checks that support our galaxy modeling results.  Various
systematics, such as the smaller number and larger uncertainties of
measurements at lower luminosities as well as internal reddening, are
discussed in the context of their effect on the \rl\ relationship.

\acknowledgements 

We thank the anonymous referee for helpful comments and suggestions
which improved the presentation of this manuscript.  We also thank
Chien Peng for his excellent program Galfit, which enabled us to carry
out this study, and for helpful conversations regarding the galaxy
fitting.  This work is based on observations with the NASA/ESA {\it
Hubble Space Telescope}.  We are grateful for support of this work
through grants {\it HST} GO-9851, GO-10516, and GO-10833 from the
Space Telescope Science Institute, which is operated by the
Association of Universities for Research in Astronomy, Inc., under
NASA contract NAS5-26555, and by the NSF through grant AST-0604066 to
The Ohio State University.  M.B. gratefully acknowledges support from
the NSF through grant AST-0548198 to the University of California,
Irvine.  This research has made use of the NASA/IPAC Extragalactic
Database (NED) which is operated by the Jet Propulsion Laboratory,
California Institute of Technology, under contract with the National
Aeronautics and Space Administration and the SIMBAD database, operated
at CDS, Strasbourg, France.

\clearpage

\begin{figure}
\epsscale{0.85}
\plotone{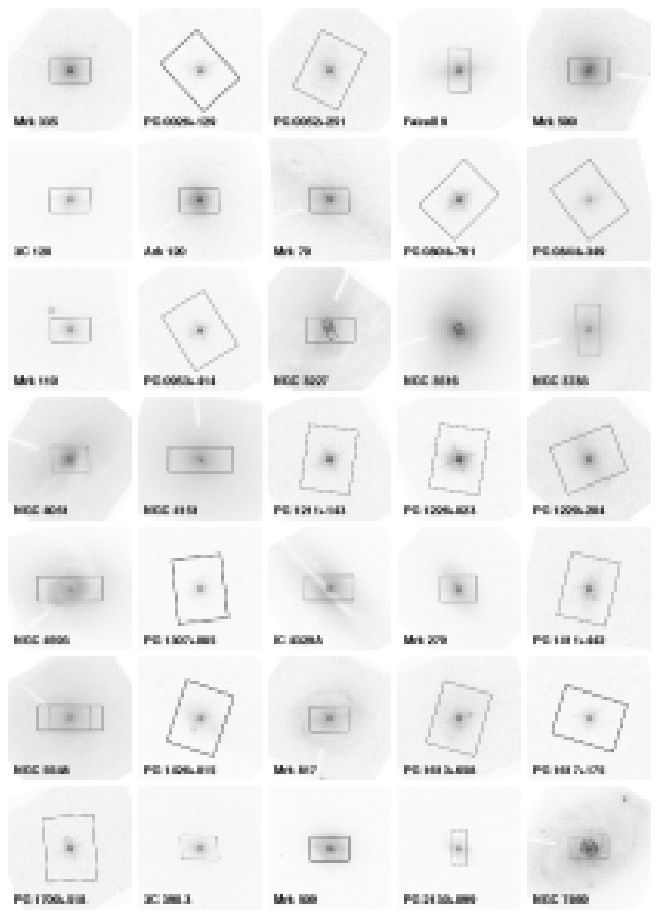}
\caption{The final stacked images for the full sample of 35
         reverberation-mapped AGNs with \hst\ imaging.  The
         ground-based spectroscopic monitoring apertures are overlaid.
         Each image is 25\arcsec~$\times$~25\arcsec, with north up and
         east to the left.}
\end{figure}

\begin{figure}
\epsscale{0.6}
\plotone{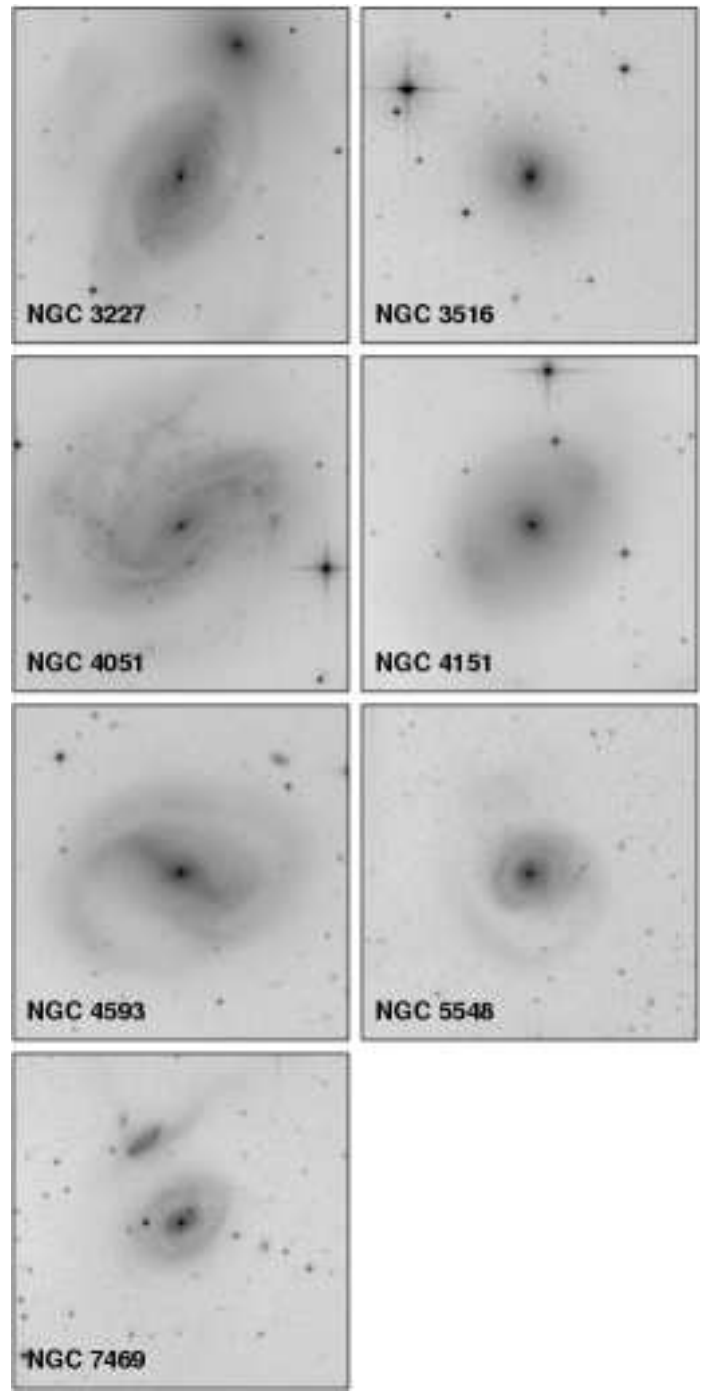}
\caption{$V$-band images of the seven NGC objects in the
         reverberation-mapped sample that are visible from MDM
         Observatory.  Each image is 5\arcmin~$\times$~5\arcmin, with
         north up and east to the left.}
\end{figure}

\begin{figure}
\epsscale{1}
\plotone{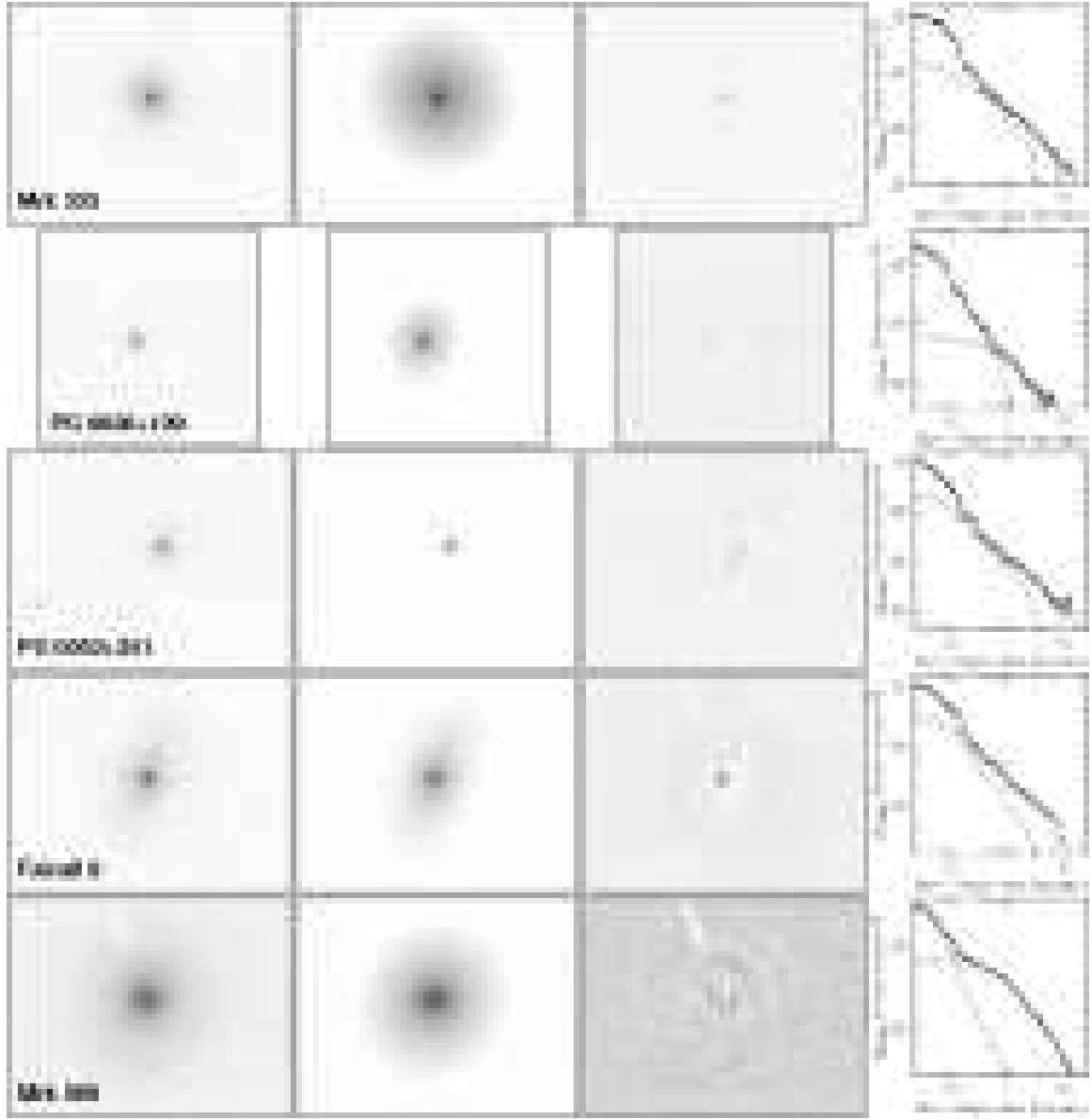}
\caption{From left to right, each row shows the following: the \hst\
         image of an AGN and its host galaxy; the best-fit to the
         galaxy+AGN surface brightness profiles from Galfit; the
         residuals of the fit; isophotal analysis of the image and
         models, with the data points measured from the sky-subtracted
         \hst\ image, the solid line from the total host-galaxy model
         image (which is convolved with the PSF for comparison with
         the observed image), and the dashed line from the PSF model
         for the AGN component.}
\end{figure}

\begin{figure}
\figurenum{3}
\plotone{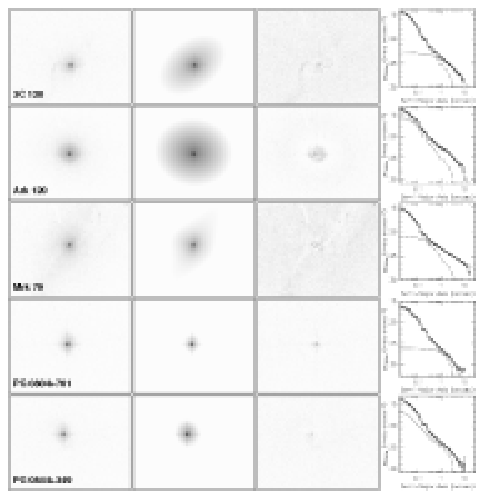}
\caption{{\it Continued.}}
\end{figure}

\begin{figure}
\figurenum{3}
\plotone{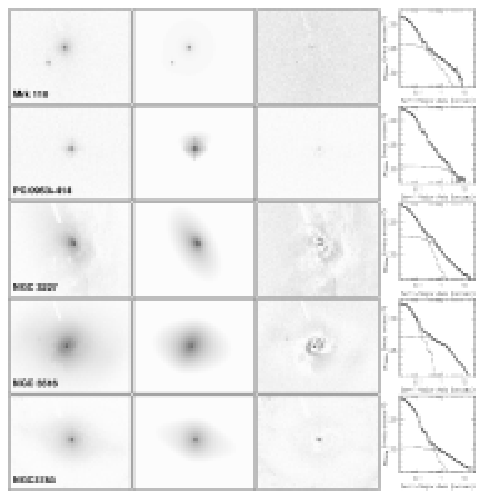}
\caption{{\it Continued.}}
\end{figure}

\begin{figure}
\figurenum{3}
\plotone{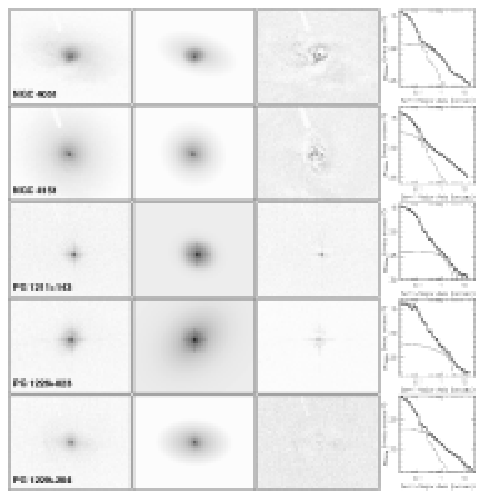}
\caption{{\it Continued.}}
\end{figure}

\begin{figure}
\figurenum{3}
\plotone{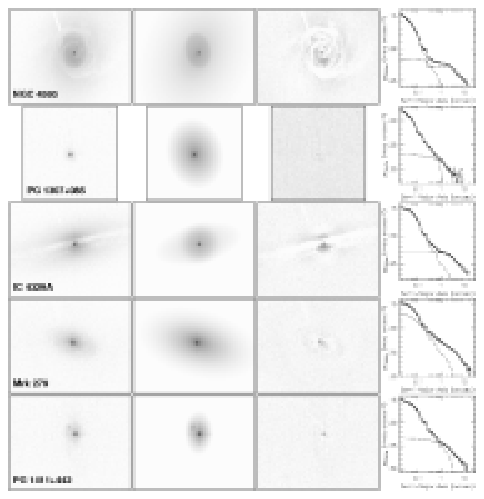}
\caption{{\it Continued.}}
\end{figure}

\begin{figure}
\figurenum{3}
\plotone{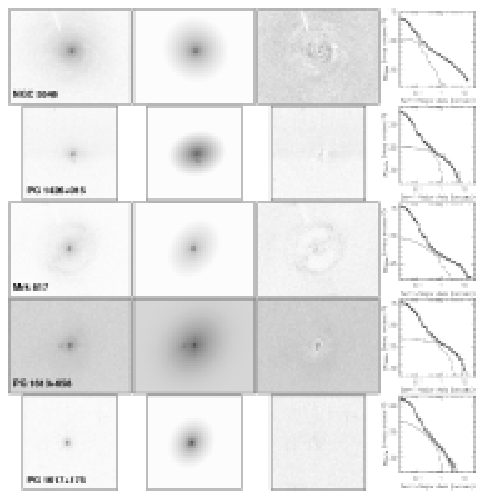}
\caption{{\it Continued.}}
\end{figure}

\begin{figure}
\figurenum{3}
\plotone{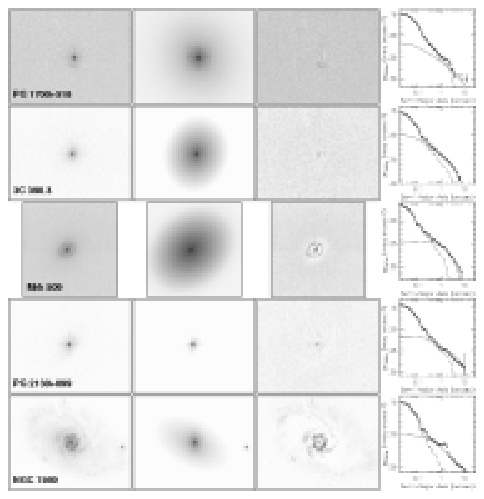}
\caption{{\it Continued.}}
\end{figure}

\begin{figure}
\plotone{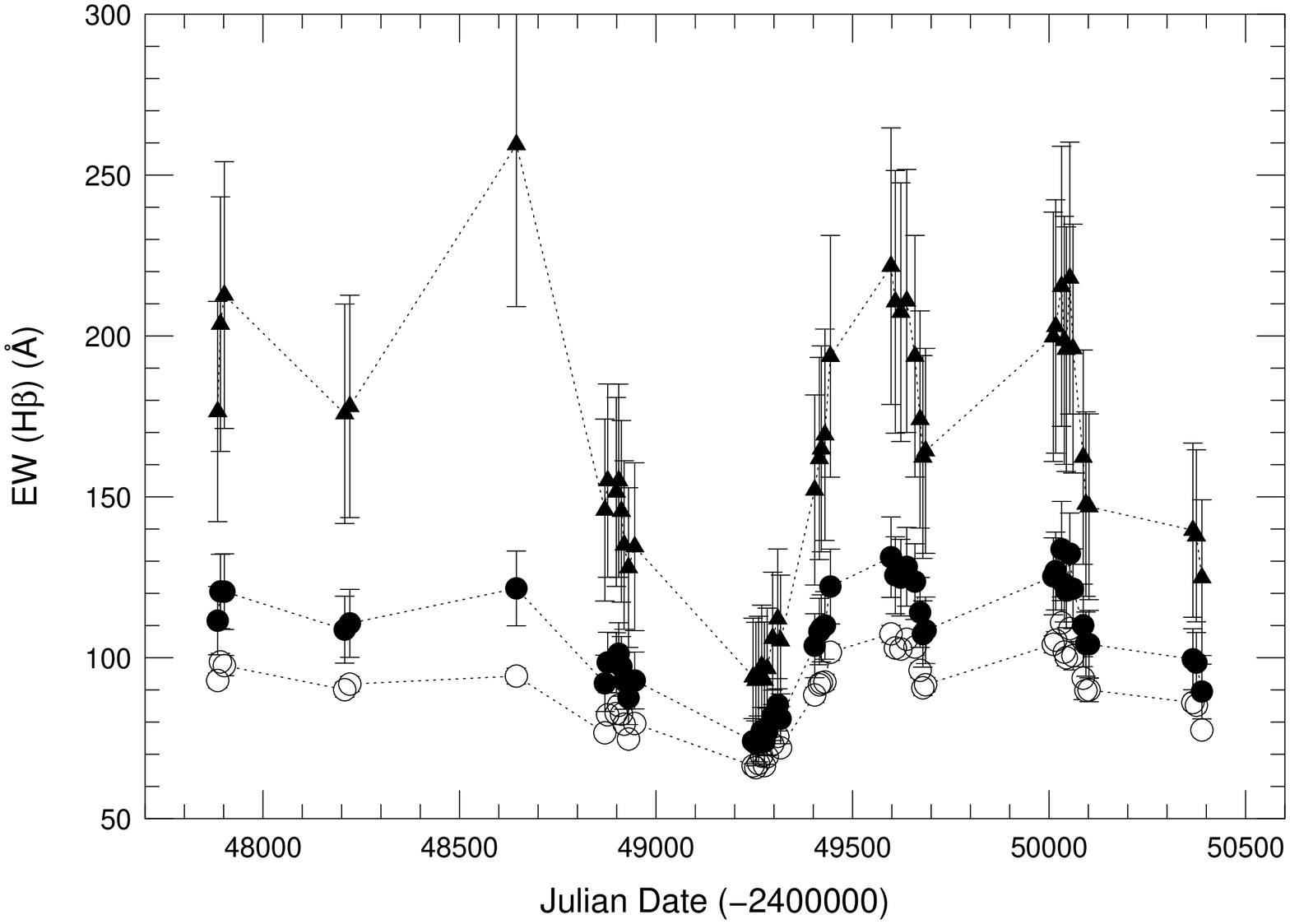}
\caption{Equivalent width consistency check for 3C\,120.  The
         equivalent width of the H$\beta$ emission line is shown as a
         function of time for three cases: with no correction of the
         continuum flux density for host-galaxy contamination (open
         circles); with corrections for the host galaxy using the
         value from \citet{bentz06a} (triangles); and with corrections
         for the host galaxy using the revised value presented here in
         Table 7 (filled circles). The larger host-galaxy correction
         from \citeauthor{bentz06a} leads to equivalents widths that
         are unacceptably large and variable.  The new, more
         conservative value for the starlight contribution corrects
         this problem so that the equivalent width values and
         variations are in keeping with expectations from
         photoionization physics.}
\end{figure}

\begin{figure}
\plotone{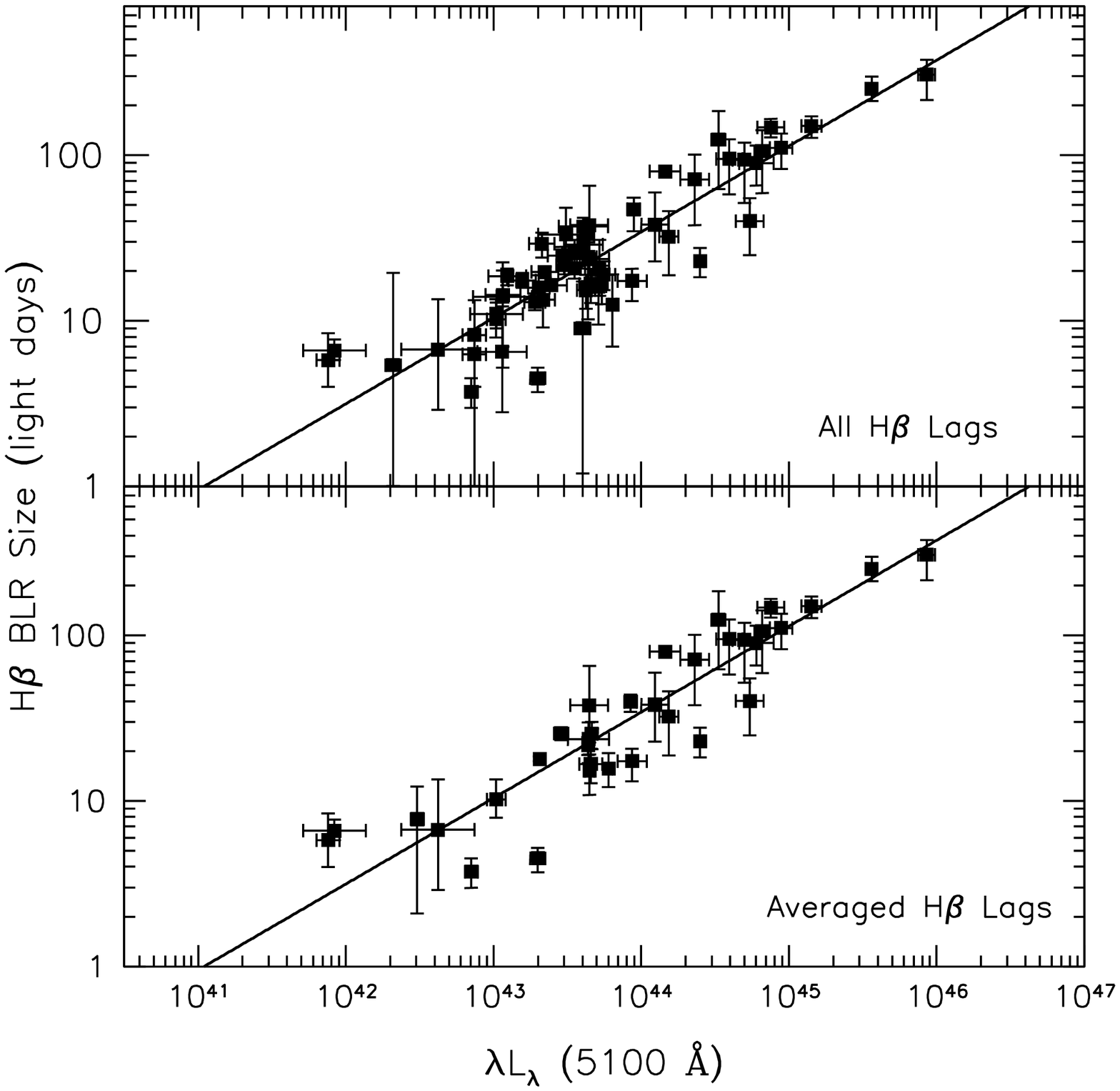}
\caption{The H$\beta$ \rl\ relationship after correcting the AGN
         luminosities for the contribution from host-galaxy starlight.
         The top panel shows each separate measurement as a single
         data point, and the bottom panel shows the weighted mean of
         multiple measurements for any individual object.  The solid
         lines are the best fit to the relationship (listed in bold
         face in Table~9), which has a slope of $\alpha =
         0.519^{+0.063}_{-0.066}$.}
\end{figure}

\clearpage

\begin{deluxetable}{lllcccl}
\tablecolumns{7}
\tablewidth{0pt}
\tabletypesize{\scriptsize}
\tablecaption{Object List}
\tablehead{
\colhead{Object} &
\colhead{$\alpha_{\rm 2000}$} &
\colhead{$\delta_{\rm 2000}$} &
\colhead{$z$} &
\colhead{$D_{\rm L}$\tablenotemark{a}} &
\colhead{$A_{\rm B}$\tablenotemark{b}} &
\colhead{Alternate} \\
\colhead{} &
\colhead{(hr min sec)} &
\colhead{(\arcdeg\ \arcmin\ \arcsec)} &
\colhead{} &
\colhead{Mpc} &
\colhead{(mag)} &
\colhead{Name}}

\startdata
Mrk\,335     & 00\,06\,19.521  & +20\,12\,10.49  & 0.02579 & 112.6  & 0.153 & PG\,0003+199 \\
PG\,0026+129 & 00\,29\,13.6    & +13\,16\,03     & 0.14200 & 671.7  & 0.307 & \\
PG\,0052+251 & 00\,54\,52.1    & +25\,25\,38     & 0.15500 & 740.0  & 0.205 & \\
Fairall\,9   & 01\,23\,45.780  & -58\,48\,20.50  & 0.04702 & 208.6  & 0.116 & \\
Mrk\,590     & 02\,14\,33.562  & -00\,46\,00.09  & 0.02639 & 115.3  & 0.161 & NGC\,863 \\
3C\,120	     & 04\,33\,11.0955 & +05\,21\,15.620 & 0.03301 & 145.0  & 1.283 & Mrk\,1506 \\
Ark\,120     & 05\,16\,11.421  & -00\,08\,59.38  & 0.03271 & 141.8  & 0.554 & Mrk\,1095 \\
Mrk\,79      & 07\,42\,32.797  & +49\,48\,34.75  & 0.02219 & 96.7   & 0.305 & \\
PG\,0804+761 & 08\,10\,58.600  & +76\,02\,42.00  & 0.10000 & 460.5  & 0.150 & \\
PG\,0844+349 & 08\,47\,42.4    & +34\,45\,04     & 0.06400 & 287.4  & 0.159 & \\
Mrk\,110     & 09\,25\,12.870  & +52\,17\,10.52  & 0.03529 & 155.2  & 0.056 & \\
PG\,0953+414 & 09\,56\,52.4    & +41\,15\,22     & 0.23410 & 1172.1 & 0.054 & \\
NGC\,3227    & 10\,23\,30.5790 & +19\,51\,54.180 & 0.00386 & 23.6   & 0.098 & \\
NGC\,3516    & 11\,06\,47.490  & +72\,34\,06.88  & 0.00884 & 38.1   & 0.183 & \\
NGC\,3783    & 11\,39\,01.72   & -37\,44\,18.9   & 0.00973 & 42.0   & 0.514 & \\
NGC\,4051    & 12\,03\,09.614  & +44\,31\,52.80  & 0.00234 & 15.2   & 0.056 & \\
NGC\,4151    & 12\,10\,32.579  & +39\,24\,20.63  & 0.00332 & 14.3   & 0.119 & \\
PG\,1211+143 & 12\,14\,17.7    & +14\,03\,12.6   & 0.08090 & 367.6  & 0.150 & \\
PG\,1226+023 & 12\,29\,06.6997 & +02\,03\,08.598 & 0.15834 & 757.5  & 0.089 & 3C\,273 \\
PG\,1229+204 & 12\,32\,03.605  & +20\,09\,29.21  & 0.06301 & 282.8  & 0.117 & Mrk\,771 \& Ton\,1542 \\
NGC\,4593    & 12\,39\,39.425  & -05\,20\,39.34  & 0.00900 & 38.8   & 0.106 & Mrk\,1330\\
PG\,1307+085 & 13\,09\,47.0    & +08\,19\,48.9   & 0.15500 & 739.2  & 0.145 & \\
IC\,4329A    & 13\,49\,19.26   & -30\,18\,34.0   & 0.01605 & 69.6   & 0.255 & \\
Mrk\,279     & 13\,53\,03.447  & +69\,18\,29.57  & 0.03045 & 133.5  & 0.068 & \\
PG\,1411+442 & 14\,13\,48.3    & +44\,00\,14     & 0.08960 & 409.7  & 0.036 & \\
NGC\,5548    & 14\,17\,59.534  & +25\,08\,12.44  & 0.01718 & 74.5   & 0.088 & \\
PG\,1426+015 & 14\,29\,06.588  & +01\,17\,06.48  & 0.08647 & 394.4  & 0.137 & \\
Mrk\,817     & 14\,36\,22.068  & +58\,47\,39.38  & 0.03146 & 137.9  & 0.029 & PG\,1434+590 \\
PG\,1613+658 & 16\,13\,57.179  & +65\,43\,09.58  & 0.12900 & 605.6  & 0.114 & Mrk\,876 \\
PG\,1617+175 & 16\,20\,11.288  & +17\,24\,27.70  & 0.11244 & 521.9  & 0.180 & Mrk\,877 \\
PG\,1700+518 & 17\,01\,24.800  & +51\,49\,20.00  & 0.29200 & 1509.6 & 0.151 & \\
3C\,390.3    & 18\,42\,08.9899 & +79\,46\,17.127 & 0.05610 & 250.5  & 0.308 & \\
Mrk\,509     & 20\,44\,09.738  & -10\,43\,24.54  & 0.03440 & 151.2  & 0.248 & \\
PG\,2130+099 & 21\,32\,27.813  & +10\,08\,19.46  & 0.06298 & 282.6  & 0.192 & II\,Zw\,136 \& Mrk\,1513\\
NGC\,7469    & 23\,03\,15.623  & +08\,52\,26.39  & 0.01632 & 70.8   & 0.297 & Mrk\,1514 \\

\enddata

\tablenotetext{a}{Distances were calculated from the
                  redshifts of the objects, except for NGC\,3227 --
                  where we use the distance measured by surface
                  brightness fluctuations to
                  NGC\,3226 \citep{blakeslee01}, with which NGC\,3227
                  is interacting -- and NGC\,4051 -- where we use the
                  average Tully-Fisher distance reported by
                  \citet{russell03}.}

\tablenotetext{b}{Values are from \citet{schlegel98}.}

\end{deluxetable}

\clearpage

\begin{deluxetable}{lccccc}
\tablecolumns{6}
\tablewidth{0pt}
\tabletypesize{\scriptsize}
\tablecaption{\hst\ Observation Log}
\tablehead{
\colhead{Object} &
\colhead{Observational} &
\colhead{Date Observed} &
\colhead{Beginning UTC} &
\colhead{Total Exposure} &
\colhead{Dataset}\\
\colhead{} &
\colhead{Setup}&
\colhead{(yyyy--mm--dd)} &
\colhead{(hh:mm)} &
\colhead{Time (s)} &
\colhead{}}

\startdata
Mrk\,335     & ACS,HRC,F550M & 2006-08-24 & 08:26 & 2040 & J9MU010 \\
PG\,0026+129 & WFPC2,F547M   & 2007-06-06 & 21:39 & 1445 & U9MU520 \\
PG\,0052+251 & ACS,HRC,F550M & 2006-08-28 & 09:59 & 2040 & J9MU030 \\
Fairall\,9   & ACS,HRC,F550M & 2003-08-22 & 00:44 & 1020 & J8SC040 \\
Mrk\,590     & ACS,HRC,F550M & 2003-12-18 & 02:27 & 1020 & J8SC050 \\
3C\,120      & ACS,HRC,F550M & 2003-12-05 & 05:48 & 1020 & J8SC060 \\
Akn\,120     & ACS,HRC,F550M & 2006-10-30 & 18:36 & 2040 & J9MU540 \\   
Mrk\,79      & ACS,HRC,F550M & 2006-11-08 & 18:33 & 2040 & J9MU050 \\
PG\,0804+761 & ACS,HRC,F550M & 2006-09-20 & 23:09 & 2040 & J9MU060 \\
PG\,0844+349 & ACS,HRC,F550M & 2004-05-10 & 20:11 & 1020 & J8SC100 \\
Mrk\,110     & ACS,HRC,F550M & 2004-05-28 & 17:34 & 1020 & J8SC110 \\
PG\,0953+414 & ACS,HRC,F550M & 2006-10-25 & 18:51 & 2040 & J9MU070 \\
NGC\,3227    & ACS,HRC,F550M & 2004-03-20 & 04:28 & 1020 & J8SC130 \\
NGC\,3516    & ACS,HRC,F550M & 2005-12-19 & 01:03 & 2220 & J9DQ010 \\
NGC\,3783    & ACS,HRC,F550M & 2003-11-15 & 00:11 & 1020 & J8SC150 \\
NGC\,4051    & ACS,HRC,F550M & 2004-02-16 & 01:49 & 1020 & J8SC160 \\
NGC\,4151    & ACS,HRC,F550M & 2004-03-28 & 14:25 & 1020 & J8SC170 \\
PG\,1211+143 & ACS,HRC,F550M & 2006-11-28 & 02:30 & 2040 & J9MU080 \\
PG\,1226+023 & ACS,HRC,F550M & 2007-01-17 & 12:40 & 2040 & J9MU090 \\
PG\,1229+204 & ACS,HRC,F550M & 2006-11-20 & 02:41 & 2040 & J9MU100 \\
NGC\,4593    & ACS,HRC,F550M & 2006-01-30 & 21:05 & 2220 & J9DQ020 \\
PG\,1307+085 & WFPC2,F547M   & 2007-03-21 & 14:36 & 1445 & U9MU110 \\
IC\,4329A    & ACS,HRC,F550M & 2006-02-22 & 00:03 & 2220 & J9DQ040 \\
Mrk\,279     & ACS,HRC,F550M & 2003-12-07 & 03:54 & 1020 & J8SC240 \\ 
PG\,1411+442 & ACS,HRC,F550M & 2006-11-10 & 23:52 & 2040 & J9MU120 \\
NGC\,5548    & ACS,HRC,F550M & 2004-04-07 & 01:53 & 1020 & J8SC270 \\
PG\,1426+015 & WFPC2,F547M   & 2007-03-20 & 16:18 & 1445 & U9MU130 \\
Mrk\,817     & ACS,HRC,F550M & 2003-12-08 & 18:08 & 1020 & J8SC290 \\
PG\,1613+658 & ACS,HRC,F550M & 2006-11-12 & 04:44 & 2040 & J9MU140 \\
PG1617\,175  & WFPC2,F547M   & 2007-03-19 & 17:59 & 1445 & U9MU150 \\
PG\,1700+518 & ACS,HRC,F550M & 2006-11-16 & 07:51 & 2040 & J9MU160 \\
3C\,390.3    & ACS,HRC,F550M & 2004-03-31 & 06:56 & 1020 & J8SC340 \\
Mrk\,509     & WFPC2,F547M   & 2007-04-01 & 22:50 & 1445 & U9MU170 \\
PG\,2130+099 & ACS,HRC,F550M & 2003-10-21 & 06:47 & 1020 & J8SC360 \\
NGC\,7469    & ACS,HRC,F550M & 2006-07-09 & 22:00 & 2220 & J9DQ030 \\ 
\enddata                            

\end{deluxetable}

\begin{deluxetable}{lcccccc}
\tablecolumns{7}
\tablewidth{0pt}
\tablecaption{MDM Observation Log}
\tablehead{
\colhead{} &
\colhead{} &
\colhead{} &
\multicolumn{3}{c}{Exposure Times} &
\colhead{}\\
\cline{4-6}
\colhead{Object} &
\colhead{Date Observed} &
\colhead{Airmass} &
\colhead{$B$} &
\colhead{$V$} &
\colhead{$R$} &
\colhead{Seeing}\\
\colhead{} &
\colhead{(yyyy--mm--dd)} &
\colhead{(sec z)} &
\colhead{(s)} &
\colhead{(s)} &
\colhead{(s)} &
\colhead{($\arcsec$)}}

\startdata
NGC\,3227    & 2003-02-08 & $1.023-1.135$ & 1800 & 1680 & 795  & 1.77 \\
NGC\,3516    & 2003-04-25 & $1.317-1.392$ & 1900 & 845  & 1500 & 1.85 \\
NGC\,4051    & 2003-02-08 & $1.048-1.111$ & 1250 & 795  & 690  & 1.71 \\
NGC\,4151    & 2003-02-09 & $1.008-1.050$ & 1470 & 1200 & 1220 & 2.07 \\
NGC\,4593    & 2003-02-09 & $1.298-1.576$ & 1650 & 1860 & 1380 & 2.58 \\
NGC\,5548    & 2003-04-25 & $1.018-1.308$ & 4000 & 2500 & 1080 & 1.63 \\
NGC\,7469    & 2003-09-26 & $1.087-1.106$ & 1440 & 1260 & 1560 & 1.69 \\
\enddata                            

\end{deluxetable}

\clearpage

\LongTables
\begin{deluxetable}{lcccccl}
\tablecolumns{7}
\tablewidth{0pt}
\tablecaption{Details of Galaxy Fits}
\tablehead{
\colhead{Object} &
\colhead{Sky} &
\colhead{$m_{stmag}$\tablenotemark{a}} &
\colhead{$R_{\rm e}$} &
\colhead{$n$} &
\colhead{$b/a$} &
\colhead{Note} \\
\colhead{} &
\colhead{(counts)} &
\colhead{} &
\colhead{(kpc)} &
\colhead{} &
\colhead{} &
\colhead{}}

\startdata
Mrk\,335 &	10.0	& 14.9	&  	        &		&		&	PSF	\\
	 &		& 16.3	&  0.51  	&	6.61	&	0.84	&	Bulge	\\
	 &		& 15.8	&  1.49  	&	1.00	&	0.95	&	Disk	\\
PG\,0026+129 &	2.8	& 16.1	&        	&		&		&	PSF	\\
	&		& 17.0	&  0.16  	&	2.58	&	0.52	&	Add'l PSF \\
	&		& 17.3	&  7.21  	&	1.72	&	0.86	&	Bulge	\\
PG\,0052+251 &	5.7	& 15.6	&        	&		&		&	PSF	\\
	&		& 18.0	&  0.17  	&	3.12	&	0.34	&	Bulge	\\
	&		& 17.2	&  8.24  	&	1.00	&	0.69	&	Disk	\\
	&		& 18.8	&  3.77  	&	5.47	&	0.78	&	Field galaxy	\\
Fairall\,9 &	7.0	& 15.1	&        	&		&		&	PSF	\\
	&		& 15.3	&  0.49  	&	5.61	&	0.94	&	Bulge	\\
	&		& 15.3	&  2.92  	&	1.00	&	0.44	&	Disk	\\
Mrk\,590 &	6.8	& 17.9	&        	&		&		&	PSF	\\
	&		& 16.3	&  0.44  	&	1.22	&	0.62	&	Inner bulge	\\
	&		& 15.8	&  0.75  	&	0.59	&	0.96	&	Bulge	\\
	&		& 14.2	&  3.14  	&	1.00	&	0.91	&	Disk	\\
3C\,120	&	8.0	& 14.9	&        	&		&		&	PSF	\\
	&		& 18.3	&  0.04  	&	3.67	&	0.10	&	Add'l PSF	\\
	&		& 17.6	&  0.66  	&	1.10	&	0.89	&	Bulge	\\
	&		& 16.1	&  3.11  	&	1.00	&	0.65	&	Disk	\\
Ark\,120 &	12.3	& 14.7	&        	&		&		&	PSF	\\
	&		& 15.0	&  0.05  	&	3.62	&	0.84	&	Bulge	\\
	&		& 14.9	&  1.85  	&	1.00	&	0.87	&	Disk	\\
Mrk\,79	&	13.7	& 15.7	&        	&		&		&	PSF	\\
	&		& 16.1	&  0.85  	&	2.79	&	0.67	&	Bulge	\\
	&		& 16.6	&  1.98  	&	1.00	&	0.66	&	Disk	\\
	&		& 14.7	&  8.89  	&	0.55	&	0.24	&	Bar	\\
PG\,0804+761 &	10.7	& 15.0	&        	&		&		&	PSF	\\
	&		& 14.7	&  0.08  	&	1.22	&	0.73	&	Add'l PSF	\\
	&		& 16.8	&  3.33  	&	1.00	&	0.74	&	Bulge	\\
PG\,0844+349 &	6.0	& 14.6	&        	&		&		&	PSF	\\
	&		& 16.9	&  0.04  	&	2.28	&	0.12	&	Bulge	\\
	&		& 16.7	&  2.86  	&	1.00	&	0.75	&	Disk	\\
Mrk\,110 &	3.7	& 16.1	&        	&		&		&	PSF	\\
	&		& 18.2	&  0.25  	&	1.35	&	0.85	&	Bulge	\\
	&		& 16.5	&  1.73  	&	1.00	&	0.93	&	Disk	\\
	&		& 16.6	&        	&		&		&	Star	\\
NGC\,3227 &	14.2	& 15.2	&        	&		&		&	PSF	\\
	&		& 15.4	&  0.01  	&	1.51	&	0.92	&	Add'l PSF	\\
	&		& 14.9	&  0.06  	&	1.08	&	0.68	&	Inner bulge	\\
	&		& 12.8	&  1.19  	&	2.14	&	0.50	&	Bulge	\\
	&		& 13.8	&  4.66  	&	1.00	&	0.47	&	Disk	\\
NGC\,3516 &	12.0	& 15.2	&        	&		&		&	PSF	\\
	&		& 13.4	&  0.38  	&	1.24	&	0.77	&	Inner bulge	\\
	&		& 13.0	&  1.74  	&	0.96	&	0.60	&	Bulge	\\
	&		& 14.4	&  4.53  	&	1.00	&	0.52	&	Disk	\\
PG\,0953+414 &	10.0	& 14.9	&        	&		&		&	PSF	\\
	&		& 17.7	&  28.82 	&	1.39	&	0.55	&	Bulge	\\
NGC\,3783 &	12.0	& 14.2	&        	&		&		&	PSF	\\
	&		& 14.7	&  0.49  	&	1.09	&	0.92	&	Bulge	\\
	&		& 15.0	&  1.95  	&	0.33	&	0.29	&	Bar	\\
	&		& 12.0	&  6.02  	&	1.00	&	0.83	&	Disk	\\
NGC\,4051 &	8.0	& 14.8	&        	&		&		&	PSF	\\
	&		& 15.1	&  0.03  	&	1.07	&	0.85	&	Inner bulge	\\
	&		& 15.1	&  0.07  	&	0.31	&	0.71	&	Inner bulge	\\
	&		& 12.8	&  0.86  	&	1.80	&	0.49	&	Bulge	\\
	&		& 12.4	&  4.24  	&	1.00	&	0.67	&	Disk	\\
NGC\,4151 &	0.2	& 14.5	&        	&		&		&	PSF	\\
	&		& 14.4	&  0.07  	&	4.29	&	0.54	&	Inner bulge	\\
	&		& 14.0	&  0.14  	&	0.71	&	0.96	&	Inner bulge	\\
	&		& 12.0	&  0.73  	&	0.81	&	0.95	&	Bulge	\\
	&		& 13.0	&  3.77  	&	1.00	&	0.69	&	Disk	\\
PG\,1211+143 &	17.0	& 22.5	&        	&		&		&	PSF	\\
	&		& 14.9	&  0.06  	&	0.08	&	0.63	&	Add'l PSF	\\
	&		& 17.3	&  2.59  	&	1.00	&	0.84	&	Bulge	\\
PG\,1226+023 &	6.5	& 15.4	&        	&		&		&	PSF	\\
	&		& 13.2	&  0.14  	&	0.22	&	0.29	&	Add'l PSF	\\
	&		& 15.6	&  11.71 	&	2.50	&	0.75	&	Bulge	\\
PG\,1229+204 &	15.6	& 16.7	&        	&		&		&	PSF	\\
	&		& 20.2	&  0.23  	&	0.82	&	0.69	&	Bar?	\\
	&		& 17.3	&  0.99  	&	1.15	&	0.87	&	Bulge	\\
	&		& 15.8	&  7.84  	&	1.00	&	0.62	&	Disk	\\
NGC\,4593 &	15.0	& 15.3	&        	&		&		&	PSF	\\
	&		& 15.1	&  0.52  	&	0.09	&	0.71	&	Inner bulge	\\
	&		& 12.3	&  2.83  	&	1.94	&	0.68	&	Bulge	\\
	&		& 13.5	&  9.68  	&	1.00	&	0.52	&	Disk	\\
PG\,1307+085 &	1.3	& 15.6	&        	&		&		&	PSF	\\
	&		& 18.1	&  0.19  	&	0.43	&	0.05	&	Add'l PSF	\\
	&		& 17.6	&  8.74  	&	1.25	&	0.80	&	Bulge	\\
IC\,4329A &	15.0	& 14.6	&        	&		&		&	PSF	\\
	&		& 13.8	&  0.01  	&	0.31	&	0.87	&	Add'l PSF	\\
	&		& 13.7	&  0.77  	&	0.39	&	0.96	&	Inner bulge	\\
	&		& 12.6	&  3.36  	&	0.50	&	0.43	&	Bulge	\\
	&		& 15.8	&  8.74  	&	1.00	&	0.41	&	Disk	\\
Mrk\,279 &	6.0	& 15.0	&        	&		&		&	PSF	\\
	&		& 16.2	&  0.05  	&	5.94	&	0.65	&	Inner bulge	\\
	&		& 16.2	&  0.97  	&	1.88	&	0.60	&	Bulge	\\
	&		& 15.2	&  3.14  	&	1.00	&	0.54	&	Disk	\\
PG\,1411+442 &	9.3	& 15.7	&        	&		&		&	PSF	\\
	&		& 15.8	&  0.05  	&	1.00	&	0.48	&	Add'l PSF	\\
	&		& 16.8	&  5.05  	&	1.71	&	0.58	&	Bulge	\\
	&		& 18.8	&  1.87  	&	2.23	&	0.64	&	Field galaxy	\\
NGC\,5548 &	0.2	& 16.7	&        	&		&		&	PSF	\\
	&		& 14.7	&  1.18  	&	4.36	&	0.86	&	Inner bulge	\\
	&		& 13.8	&  2.95  	&	1.39	&	0.90	&	Bulge	\\
	&		& 15.5	&  11.51 	&	1.00	&	0.85	&	Disk	\\
PG\,1426+015 &	1.8	& 15.2	&        	&		&		&	PSF	\\
	&		& 16.3	&  4.74  	&	1.63	&	0.77	&	Bulge	\\
	&		& 19.8	&  0.91  	&	3.14	&	0.56	&	Field galaxy	\\
	&		& 21.7	&  0.15  	&	1.98	&	0.50	&	Field galaxy	\\
Mrk\,817 &	4.7	& 15.1	&        	&		&		&	PSF	\\
	&		& 17.7	&  0.28  	&	2.44	&	0.81	&	Bulge	\\
	&		& 14.4	&  4.32  	&	1.00	&	0.74	&	Disk	\\
PG\,1613+658 &	11.0	& 15.2	&        	&		&		&	PSF	\\
	&		& 16.3	&  6.75  	&	1.35	&	0.80	&	Bulge	\\
	&		& 19.1	&  1.79  	&	3.64	&	0.60	&	Field galaxy	\\
PG\,1617+175 &	1.3	& 16.3	&        	&		&		&	PSF	\\
	&		& 17.4	&  0.15  	&	0.09	&	0.07	&	Add'l PSF	\\
	&		& 17.3	&  3.28  	&	5.35	&	0.84	&	Bulge	\\
PG\,1700+518 &	9.0	& 19.2	&        	&		&		&	PSF	\\
	&		& 15.3	&  0.15  	&	0.02	&	0.58	&	Add'l PSF	\\
	&		& 16.9	&  0.28  	&	0.37	&	0.65	&	Add'l PSF	\\
	&		& 17.7	&  12.06 	&	5.61	&	0.89	&	Bulge	\\
3C\,390.3 &	1.3	& 15.8	&        	&		&		&	PSF	\\
	&		& 17.0	&  0.77  	&	3.86	&	0.74	&	Bulge	\\
	&		& 16.9	&  2.54  	&	1.00	&	0.86	&	Disk	\\
Mrk\,509 &	4.2	& 14.5	&        	&		&		&	PSF	\\
	&		& 15.4	&  0.04  	&	0.03	&	0.48	&	Add'l PSF	\\
	&		& 15.0	&  1.85  	&	1.00	&	0.79	&	Bulge	\\
PG\,2130+099 &	4.0	& 14.8	&        	&		&		&	PSF	\\
	&		& 18.9	&  0.38  	&	0.56	&	0.37	&	Bulge	\\
	&		& 16.5	&  4.66  	&	1.00	&	0.55	&	Disk	\\
NGC\,7469 &	7.0	& 15.2	&        	&		&		&	PSF	\\
	&		& 15.0	&  0.37  	&	1.37	&	0.70	&	Inner bulge	\\
	&		& 13.4	&  3.68  	&	1.31	&	0.55	&	Bulge	\\
	&		& 14.6	&  6.99  	&	1.00	&	0.94	&	Disk	\\
	&		& 16.2	&       	&		&		&	Star	\\

\enddata  

\tablenotetext{a}{The magnitude of an object is computed as: 
                $m_{stmag} = -2.5 \log\
               (\frac{counts}{s}) + zpt$, where $zpt=24.457$
               for the F550M filter with ACS HRC and $zpt=21.685$
               for the F547M filter with the PC chip on WFPC2.}
\end{deluxetable}

\clearpage

\begin{deluxetable}{lcccccc}
\tablecolumns{7}
\tablewidth{0pt}
\tablecaption{Global Galaxy Parameters}
\tablehead{
\colhead{Object} &
\colhead{$\lambda L_{\lambda,\rm gal}{\rm (5100\,\AA)}$\tablenotemark{a}} &
\multicolumn{2}{c}{$B/T$} &
\colhead{} &
\colhead{Morphological} &
\colhead{Flag\tablenotemark{c}}\\
\cline{3-4}
\colhead{} &
\colhead{($10^{44}$ ergs s$^{-1}$)} &
\colhead{1} &
\colhead{2\tablenotemark{b}} &
\colhead{} &
\colhead{Classification} &
\colhead{}}

\startdata

Mrk\,335     &  0.37	&	0.37	&		&&	S0/a		&	        \\
PG\,0026+129 &  2.3	&	1.00	&		&&		E1	&	*       \\
PG\,0052+251 &  4.1	&	0.32	&		&&	Sb		&	        \\
Fairall\,9   &  2.4	&	0.52	&		&&		SBa	&	*       \\
Mrk\,590     &  1.4	&	0.17	&	0.28	&&	SA(s)a		&	        \\
3C\,120      &  0.73	&	0.21	&		&&	S0		&	        \\
Akn\,120     &  2.1	&	0.49	&		&&	Sb pec		&	        \\
Mrk\,79      &  0.73	&	0.19	&	0.87	&&	SBb		&	        \\
PG\,0804+761 &  1.5	&	1.00	&		&&	 	E3	&	*       \\
PG\,0844+349 &  1.2	&	0.45	&		&&	 	Sa	&	*       \\
Mrk\,110     &  0.26	&	0.17	&		&&	 	Sc	&	*       \\
PG\,0953+414 &  3.6	&	1.00	&		&&	 	E4	&	*       \\
NGC\,3227    &  0.23	&	0.65	&	0.74	&&	SAB(s) pec	&	        \\
NGC\,3516    &  0.69	&	0.52	&	0.86	&&	(R)SB(s)	&	        \\
NGC\,3783    &  1.5	&	0.07	&	0.12	&&	(R')SB(r)a	&	        \\
NGC\,4051    &  0.16	&	0.36	&	0.45	&&	SAB(rs)bc	&	        \\
NGC\,4151    &  0.20	&	0.60	&	0.75	&&	(R')SAB(rs)ab	&	        \\
PG\,1211+143 &  0.59	&	1.00	&		&&	 	E2	&	*       \\
PG\,1226+023 &  11.3	&	1.00	&		&&	 	E3	&	*       \\
PG\,1229+204 &  1.8	&	0.20	&	0.21	&&		SBc	&	*       \\
NGC\,4593    &  0.91	&	0.71	&	0.76	&&	(R)SB(rs)b	&	        \\
PG\,1307+085 &  1.8	&	1.00	&		&&		E2	&	*       \\
IC\,4329A    &  2.4	&	0.70	&	0.96	&&	SA0		&	        \\
Mrk\,279     &  0.95	&	0.21	&	0.43	&&	S0		&	        \\
PG\,1411+442 &  1.2	&	1.00	&		&&	 	E4	&	*       \\
NGC\,5548    &  0.98	&	0.60	&	0.87	&&	(R')SA(s)0/a	&	        \\
PG\,1426+015 &  1.8	&	1.00	&		&&		E2	&	*       \\
Mrk\,817     &  1.2	&	0.05	&		&&	SBc		&	        \\
PG\,1613+658 &  4.2	&	1.00	&		&&		E2	&	*       \\
PG\,1617+175  &  1.3	&	1.00	&		&&	 	E2	&	*       \\
PG\,1700+518 &  6.7	&	1.00	&		&&	 	E1	&	*       \\
3C\,390.3    &  0.87	&	0.48	&		&&	 	Sa	&	*       \\
Mrk\,509     &  0.95	&	1.00	&		&&		E2	&	*       \\
PG\,2130+099 &  0.88	&	0.10	&		&&	(R)Sa		&	        \\
NGC\,7469    &  1.4	&	0.64	&	0.79	&&	(R')SAB(rs)a	&               \\

\enddata                            

\tablenotetext{a}{Galaxy luminosities are determined directly from the
                  model parameters that were fit to the {\it HST}
                  images.  They do not include features that were not
                  modeled, such as the nuclear starburst ring in
                  NGC\,7469, and do not include the contributions from
                  field galaxies or stars.}

\tablenotetext{b}{Bulge luminosities here include the contribution
                  from any bar and/or inner bulge component.}

\tablenotetext{c}{Morphological classifications are from NED where
                  available.  Those marked with a flag are determined
                  from the galaxy fit parameters in this work, as
                  described in the text.} 

\end{deluxetable}

\clearpage

\begin{deluxetable}{lcrrcrc}
\tablecolumns{7}
\tablewidth{0pt}
\tablecaption{Ground-Based Monitoring Apertures and Observed Fluxes}
\tablehead{
\colhead{Object} &
\colhead{Reference\tablenotemark{a}} &
\colhead{PA} &
\multicolumn{3}{c}{Aperture} &
\colhead{$f({\rm (}1+z{\rm )\,5100\,\AA})$}\\ 
\colhead{} &
\colhead{} &
\colhead{(\arcdeg)} &
\multicolumn{3}{c}{( \arcsec\ $\times$ \arcsec )} &
\colhead{($10^{-15}$ ergs s$^{-1}$ cm$^{-2}$ \AA$^{-1}$})}

\startdata

Mrk\,335     & 1  & 90.0  & 5.0  & $\times$ & 7.6   & $7.68 \pm 0.53$  \\ 
             & 1  & 90.0  & 5.0  & $\times$ & 7.6   & $8.81 \pm 0.47$  \\ 
PG\,0026+129 & 2  & 42.0  & 10.0 & $\times$ & 13.0  & $2.69 \pm 0.40$  \\
PG\,0052+251 & 2  & 153.4 & 10.0 & $\times$ & 13.0  & $2.07 \pm 0.37$  \\ 
Fairall\,9   & 3  & 0.0   & 4.0  & $\times$ & 9.0   & $5.95 \pm 0.66$  \\ 
Mrk\,590     & 1  & 90.0  & 5.0  & $\times$ & 7.6   & $7.89 \pm 0.62$  \\ 
             & 1  & 90.0  & 5.0  & $\times$ & 7.6   & $5.33 \pm 0.56$  \\ 
             & 1  & 90.0  & 5.0  & $\times$ & 7.6   & $6.37 \pm 0.45$  \\ 
             & 1  & 90.0  & 5.0  & $\times$ & 7.6   & $8.43 \pm 1.30$  \\ 
3C\,120      & 1  & 90.0  & 5.0  & $\times$ & 7.6   & $4.30 \pm 0.77$  \\ 
Akn\,120     & 1  & 90.0  & 5.0  & $\times$ & 7.6   & $10.37 \pm 0.46$ \\ 
             & 1  & 90.0  & 5.0  & $\times$ & 7.6   & $7.82 \pm 0.83$  \\ 
Mrk\,79      & 1  & 90.0  & 5.0  & $\times$ & 7.6   & $6.96 \pm 0.67$  \\ 
             & 1  & 90.0  & 5.0  & $\times$ & 7.6   & $8.49 \pm 0.86$  \\ 
             & 1  & 90.0  & 5.0  & $\times$ & 7.6   & $7.40 \pm 0.72$  \\ 
PG\,0804+761 & 2  & 315.6 & 10.0 & $\times$ & 13.0  & $5.48  \pm 1.00$ \\ 
PG\,0844+349 & 2  & 36.8  & 10.0 & $\times$ & 13.0  & $3.71  \pm 0.38$ \\ 
Mrk\,110     & 1  & 90.0  & 5.0  & $\times$ & 7.6   & $3.45  \pm 0.36$ \\ 
             & 1  & 90.0  & 5.0  & $\times$ & 7.6   & $3.96  \pm 0.51$ \\ 
             & 1  & 90.0  & 5.0  & $\times$ & 7.6   & $2.64  \pm 0.86$ \\ 
PG\,0953+414 & 2  & 31.7  & 10.0 & $\times$ & 13.0  & $1.56  \pm 0.21$ \\ 
NGC\,3227    & 4  & 25.0  & 1.5  & $\times$ & 4.0   & $23.46 \pm 3.70$ \\ 
             & 5  & 90.0  & 5.0  & $\times$ & 10.0  & $12.70 \pm 0.68$ \\ 
NGC\,3516    & 6  & 25.0  & 1.5  & $\times$ & 2.0   & $7.83  \pm 2.35$ \\ 
NGC\,3783    & 7  & 0.0   & 5.0  & $\times$ & 10.0  & $11.38 \pm 0.95$ \\ 
NGC\,4051    & 8  & 90.0  & 5.0  & $\times$ & 7.5   & $13.38 \pm 0.92$ \\ 
NGC\,4151    & 9  & 90.0  & 5.0  & $\times$ & 12.75 & $23.8  \pm 3.0$  \\ 
PG\,1211+143 & 2  & 352.2 & 10.0 & $\times$ & 13.0  & $5.66  \pm 0.92$ \\
PG\,1226+023 & 2  & 171.2 & 10.0 & $\times$ & 13.0  & $21.30 \pm 2.60$ \\ 
PG\,1229+204 & 2  & 291.5 & 10.0 & $\times$ & 13.0  & $2.15  \pm 0.23$ \\ 
NGC\,4593    & 10 & 90.0  & 5.0  & $\times$ & 12.75 & $15.9  \pm 0.7$  \\ 
PG\,1307+085 & 2  & 186.5 & 10.0 & $\times$ & 13.0  & $1.79 \pm 0.18$  \\
IC\,4329A    & 11 & 90.0  & 5.0  & $\times$ & 10.0  & $5.79  \pm 0.73$ \\
Mrk\,279     & 12 & 90.0  & 5.0   & $\times$ & 7.5    & $6.90 \pm 0.69$ \\
PG\,1411+442 & 2  & 347.0 & 10.0  & $\times$ & 13.0   & $3.71 \pm 0.32$ \\
NGC\,5548    & 13 & 90.0  & 5.0   & $\times$ & 7.5    & $9.92 \pm 1.26$ \\
       	     & 13 & 90.0  & 5.0   & $\times$ & 7.5   & $7.25 \pm 1.00$ \\
             & 13 & 90.0  & 5.0   & $\times$ & 7.5   & $9.40 \pm 0.93$ \\
       	     & 13 & 90.0  & 5.0   & $\times$ & 7.5   & $6.72 \pm 1.17$ \\
       	     & 13 & 90.0  & 5.0   & $\times$ & 7.5   & $9.06 \pm 0.86$ \\
       	     & 13 & 90.0  & 5.0   & $\times$ & 7.5   & $9.76 \pm 1.10$ \\
       	     & 13 & 90.0  & 5.0   & $\times$ & 7.5   & $12.09 \pm 1.00$ \\
      	     & 13 & 90.0  & 5.0   & $\times$ & 7.5   & $10.56 \pm 1.64$ \\
    	     & 13 & 90.0  & 5.0   & $\times$ & 7.5   & $8.12 \pm 0.91$ \\
    	     & 13 & 90.0  & 5.0   & $\times$ & 7.5   & $13.47 \pm 1.45$ \\ 
    	     & 13 & 90.0  & 5.0   & $\times$ & 7.5   & $11.83 \pm 1.82$ \\
   	     & 13 & 90.0  & 5.0   & $\times$ & 7.5   & $6.98 \pm 1.20$ \\
	     & 13 & 90.0  & 5.0   & $\times$ & 7.5   & $7.03 \pm 0.86$ \\
	     & 14 & 90.0  & 5.0   & $\times$ & 12.75  & $6.63 \pm 0.36$ \\ 
PG\,1426+015 & 2  & 341.4 & 10.0  & $\times$ & 13.0   & $4.62 \pm 0.71$ \\
Mrk\,817     & 1  & 90.0  & 5.0   & $\times$ & 7.6    & $6.10 \pm 0.83$ \\
             & 1  & 90.0  & 5.0   & $\times$ & 7.6    & $5.00 \pm 0.49$ \\
	     & 1  & 90.0  & 5.0   & $\times$ & 7.6    & $5.01 \pm 0.27$ \\
PG\,1613+658 & 2  & 164.2 & 10.0  & $\times$ & 13.0   & $3.49 \pm 0.43$ \\
PG\,1617+175 & 2  & 253.0 & 10.0  & $\times$ & 13.0   & $1.44 \pm 0.25$ \\
PG\,1700+518 & 2  & 183.5 & 10.0  & $\times$ & 13.0   & $2.20 \pm 0.15$ \\
3C\,390.3    & 15 & 90.0  & 5.0   & $\times$ & 7.5    & $1.73 \pm 0.28$ \\
Mrk\,509     & 1  & 90.0  & 5.0   & $\times$ & 7.6    & $10.92 \pm 1.99$ \\
PG\,2130+099 & 16 & 0.0   & 3.0   & $\times$ & 6.97   & $4.63 \pm 0.23$ \\
NGC\,7469    & 17 & 90.0  & 5.0   & $\times$ & 7.5    & $13.57 \pm 0.61$\\ 
					 
\enddata

\tablecomments{Here, and throughout, observed galaxy fluxes are
               tabulated at rest-frame 5100\,\AA.}

\tablenotetext{a}{References refer to reverberation-mapping campaigns in 
                  optical wavelengths.}

\tablerefs{1. \citet{peterson98a},
	   2. \citet{kaspi00},
	   3. \citet{santoslleo97},
	   4. \citet{salamanca94},
           5. \citet{winge95},
           6. \citet{wanders93},			 
           7. \citet{stirpe94},				
           8. \citet{peterson00},			
           9. \citet{bentz06b},				    
          10. \citet{denney06},				    
          11. \citet{winge96},				    
          12. \citet{santoslleo01},			    
          13. \citet{peterson02} and references therein,    
          14. \citet{bentz07},				    
          15. \citet{dietrich98},			    
	  16. \citet{grier08},			    
	  17. \citet{collier98}.       
}

\end{deluxetable}

\clearpage

\begin{deluxetable}{lcccc}
\tablecolumns{5}
\tablewidth{0pt}
\tabletypesize{\scriptsize}
\tablecaption{Host-Galaxy Fluxes and Luminosities}
\tablehead{
\colhead{Object} &
\colhead{$f_{\rm gal} (HST)$} &
\colhead{$\underline{f_{\rm gal} (HST)}$} &
\colhead{$f_{\rm gal}{\rm ((}1+z{\rm )\,5100\,\AA)}$} &
\colhead{$\lambda L_{\lambda,{\rm gal}}{\rm (5100\,\AA)}$\tablenotemark{a}} \\
\colhead{} &
\colhead{($10^{-15}$ ergs s$^{-1}$ cm$^{-2}$ \AA$^{-1}$)} &
\colhead{$f_{\rm gal}{\rm ((}1+z{\rm )\,5100\,\AA)}$} &
\colhead{($10^{-15}$ ergs s$^{-1}$ cm$^{-2}$ \AA$^{-1}$)} &
\colhead{($10^{44}$ ergs s$^{-1}$)}
}

\startdata

Mrk\,335     & 1.88  	& 0.85	  & $ 1.60   \pm  0.15 $  &    $0.142   \pm 0.013   $  \\
PG\,0026+129 & 0.381	& 1.00	  & $ 0.379  \pm  0.035$  &    $1.46    \pm 0.13    $  \\
PG\,0052+251 & 0.713	& 0.98	  & $ 0.699  \pm  0.064$  &    $3.08    \pm 0.28    $  \\
Fairall\,9   & 3.47	& 0.88	  & $ 3.07   \pm  0.28 $  &    $0.927   \pm 0.085   $  \\
Mrk\,590     & 4.81	& 0.85	  & $ 4.10   \pm  0.38 $  &    $0.384   \pm 0.035   $  \\
3C\,120      & 0.783	& 0.82	  & $ 0.641  \pm  0.059$  &    $0.217   \pm 0.020   $  \\
Akn\,120     & 6.70	& 0.85	  & $ 5.68   \pm  0.52 $  &    $1.079   \pm 0.099   $  \\
Mrk\,79      & 1.74	& 0.84	  & $ 1.46   \pm  0.13 $  &    $0.106   \pm 0.010   $  \\
PG\,0804+761 & 0.703	& 0.97	  & $ 0.683  \pm  0.063$  &    $1.076   \pm 0.099   $  \\
PG\,0844+349 & 1.24	& 0.92	  & $ 1.14   \pm  0.11 $  &    $0.684   \pm 0.063   $  \\
Mrk\,110     & 0.786	& 0.88	  & $ 0.688  \pm  0.063$  &    $0.109   \pm 0.010   $  \\
PG\,0953+414 & 0.208	& 1.11	  & $ 0.231  \pm  0.021$  &    $2.47    \pm 0.23    $  \\
NGC\,3227\tablenotemark{b}    & 4.12	& 0.82	  & $ 3.37   \pm  0.31 $  &    $0.0124  \pm 0.0011  $  \\
             & 8.57	& 0.82	  & $ 7.01   \pm  0.65 $  &    $0.0258  \pm 0.0024  $  \\
NGC\,3516    & 4.55	& 0.82	  & $ 3.73   \pm  0.34 $  &    $0.0382  \pm 0.0035  $  \\
NGC\,3783    & 6.06	& 0.80	  & $ 4.86   \pm  0.45 $  &    $0.0776  \pm 0.0071  $  \\
NGC\,4051    & 10.1	& 0.82	  & $ 8.28   \pm  0.76 $  &    $0.0123  \pm 0.0011  $  \\
NGC\,4151    & 21.6	& 0.82	  & $ 17.6   \pm  1.6  $  &    $0.0240  \pm 0.0022  $  \\
PG\,1211+143 & 0.633	& 0.95	  & $ 0.598  \pm  0.055$  &    $0.592   \pm 0.054   $  \\
PG\,1226+023 & 1.37	& 0.98	  & $ 1.34   \pm  0.12 $  &    $5.741   \pm 0.529   $  \\
PG\,1229+204 & 1.48	& 0.92	  & $ 1.36   \pm  0.13 $  &    $0.766   \pm 0.071   $  \\
NGC\,4593    & 10.7	& 0.82	  & $ 8.85   \pm  0.82 $  &    $0.0889  \pm 0.0082  $  \\
PG\,1307+085 & 0.232	& 1.00	  & $ 0.233  \pm  0.021$  &    $0.986   \pm 0.091   $  \\
IC\,4329A    & 4.43	& 0.83	  & $ 3.67   \pm  0.34 $  &    $0.133   \pm 0.012   $  \\
Mrk\,279     & 3.49	& 0.87	  & $ 3.02   \pm  0.28 $  &    $0.355   \pm 0.033   $  \\
PG\,1411+442 & 0.826	& 0.96	  & $ 0.791  \pm  0.073$  &    $0.904   \pm 0.083   $  \\
NGC\,5548\tablenotemark{b}    & 4.63	& 0.84	  & $ 3.88   \pm  0.36 $  &    $0.143   \pm 0.013   $  \\
             & 5.51	& 0.84	  & $ 4.61   \pm  0.43 $  &    $0.169   \pm 0.016   $  \\
PG\,1426+015 & 1.19	& 0.96	  & $ 1.14   \pm  0.11 $  &    $1.29    \pm 0.12    $  \\
Mrk\,817     & 1.77	& 0.87	  & $ 1.54   \pm  0.14 $  &    $0.188   \pm 0.017   $  \\
PG\,1613+658 & 1.52	& 0.98	  & $ 1.50   \pm  0.14 $  &    $4.08    \pm 0.38    $  \\
PG\,1617+175 & 0.341	& 0.99	  & $ 0.336  \pm  0.031$  &    $0.701   \pm 0.065   $  \\
PG\,1700+518 & 0.246	& 1.41	  & $ 0.347  \pm  0.032$  &    $6.79    \pm 0.63    $  \\
3C\,390.3    & 0.945	& 0.90	  & $ 0.853  \pm  0.079$  &    $0.430   \pm 0.040   $  \\
Mrk\,509     & 2.74	& 0.92	  & $ 2.52   \pm  0.23 $  &    $0.435   \pm 0.040   $  \\
PG\,2130+099 & 0.440    & 0.92	  & $ 0.405  \pm  0.037$  &    $0.240   \pm 0.022   $  \\
NGC\,7469    & 10.2	& 0.83	  & $ 8.43   \pm  0.78 $  &    $0.327   \pm 0.030   $  \\

\enddata

\tablenotetext{a}{The galaxy luminosities presented here are measured
                  through the ground-based monitoring aperture
                  directly from the PSF-subtracted \hst\ images.  Any
                  field galaxies or stars, or additional unmodeled
                  galaxy structures, that are included in the original
                  ground-based spectroscopic aperture contribute to
                  this luminosity.}

\tablenotetext{b}{The two different entries for NGC\,3227 and
                  NGC\,5548 correspond to the two different monitoring
                  apertures that were employed during the
                  spectroscopic monitoring programs for these objects.
                  They are listed in the same order as in Table~6.}

\end{deluxetable}

\clearpage

\begin{deluxetable}{lrcc}
\tablecolumns{4}
\tablewidth{0pt}
\tablecaption{Rest-frame Time Lags and Starlight Corrected Luminosities}
\tablehead{
\colhead{Object} &
\colhead{H$\beta$ Time Lag} &
\colhead{$f_{\rm AGN}$ (($1+z$)\,5100\,\AA)} &
\colhead{$\lambda L_{\lambda,{\rm AGN}}$ (5100\,\AA)}\\
\colhead{} &
\colhead{(days)} &
\colhead{($10^{-15}$ ergs s$^{-1}$ cm$^{-2}$ \AA$^{-1}$)} &
\colhead{($10^{44}$ ergs s$^{-1}$)}
}

\startdata

Mrk\,335     &		$16.8^{+4.8 }_{-4.2 }$	&		$6.09	\pm 0.53$  &		$0.541	\pm 0.047 $  \\
 	     &    	$12.5^{+6.6 }_{-5.5 }$	&		$7.21	\pm 0.47$  &		$0.640	\pm 0.041 $  \\
	     &{\boldmath $15.7^{+3.4 }_{-4.0 }$}&{\boldmath	$6.72	\pm 0.35$} &{\boldmath	$0.603	\pm 0.031 $} \\
PG\,0026+129 &		$111.0^{+24.1}_{-28.3}$	&		$2.31	\pm 0.40$  &		$8.9	\pm 1.6 $  \\
PG\,0052+251 &		$89.8^{+24.5}_{-24.1}$	&		$1.37	\pm 0.37$  &		$6.0	\pm 1.6 $  \\
Fairall\,9   &		$17.4^{+3.2 }_{-4.3 }$	&		$2.88	\pm 0.66$  &		$0.87	\pm 0.20 $  \\
Mrk\,590     &		$20.7^{+3.5 }_{-2.7 }$	&		$3.80	\pm 0.62$  &		$0.355	\pm 0.058 $  \\
	     &    	$14.0^{+8.5 }_{-8.8 }$	&		$1.23	\pm 0.56$  &		$0.115	\pm 0.053 $  \\
	     &    	$29.2^{+4.9 }_{-5.0 }$	&		$2.27	\pm 0.45$  &		$0.212	\pm 0.043 $  \\
	     &    	$28.8^{+3.6 }_{-4.2 }$	&		$4.3	\pm 1.3$  &		$0.41	\pm 0.12 $  \\
	     &{\boldmath $25.6^{+2.0 }_{-2.3 }$}&{\boldmath	$2.44	\pm 0.30$} &{\boldmath	$0.287	\pm 0.032 $} \\
3C\,120	     &    	$38.1^{+21.3}_{-15.3}$	&		$3.66	\pm 0.77$  &		$1.24	\pm 0.26 $  \\
Ark\,120     &		$47.1^{+8.3 }_{-12.4}$	&		$4.69	\pm 0.46$  &		$0.889	\pm 0.088 $  \\
	     &    	$37.1^{+4.8 }_{-5.4 }$	&		$2.14	\pm 0.83$  &		$0.41	\pm 0.16 $  \\
	     &{\boldmath $39.7^{+3.9 }_{-5.5 }$}&{\boldmath	$4.09	\pm 0.40$} &{\boldmath	$0.847	\pm 0.081 $} \\
Mrk\,79	     &    	$9.0 ^{+8.3 }_{-7.8 }$	&		$5.50	\pm 0.67$  &		$0.401	\pm 0.049 $  \\
	     &    	$16.1^{+6.6 }_{-6.6 }$	&		$7.03	\pm 0.86$  &		$0.513	\pm 0.063 $  \\
	     &    	$16.0^{+6.4 }_{-5.8 }$	&		$5.94	\pm 0.72$  &		$0.435	\pm 0.053 $  \\
	     &{\boldmath $15.2^{+3.4 }_{-5.1 }$}&{\boldmath	$6.03	\pm 0.43$} &{\boldmath	$0.447	\pm 0.031 $} \\
PG\,0804+761 &		$146.9^{+18.8}_{-18.9}$	&		$4.8	\pm 1.0$  &		$7.6	\pm 1.6 $  \\
PG\,0844+349\tablenotemark{a} &	$32.3^{+13.7}_{-13.4}$	&	$2.57	\pm 0.38$  &		$1.54	\pm 0.23 $  \\
Mrk\,110     &		$24.3^{+5.5 }_{-8.3 }$	&		$2.77	\pm 0.36$  &		$0.439	\pm 0.058 $  \\
	     &    	$20.4^{+10.5}_{-6.3 }$	&		$3.28	\pm 0.51$  &		$0.520	\pm 0.080 $  \\
	     &      	$33.3^{+14.9}_{-10.0}$	&		$1.95	\pm 0.86$  &		$0.31	\pm 0.14 $  \\
	     &{\boldmath $25.5^{+4.2 }_{-5.6 }$}&{\boldmath	$2.83	\pm 0.28$} &{\boldmath	$0.461	\pm 0.045 $} \\
PG\,0953+414 &		$150.1^{+21.6}_{-22.6}$	&		$1.33	\pm 0.21$  &		$14.2	\pm 2.2 $  \\
NGC\,3227    &		$8.2 ^{+5.1 }_{-8.4 }$	&		$20.1	\pm 3.7$  &		$0.074	\pm 0.014 $  \\
	     &    	$5.4 ^{+14.1}_{-8.7 }$	&		$5.70	\pm 0.68$  &		$0.0209	\pm 0.0025 $  \\
	     &{\boldmath $7.8 ^{+3.5 }_{-10.2}$}&{\boldmath	$6.17	\pm 0.67$} &{\boldmath	$0.0304	\pm 0.0031 $} \\
NGC\,3516    &		$6.7 ^{+6.8 }_{-3.8 }$	&		$4.1	\pm 2.3$  &		$0.042	\pm 0.024 $  \\
NGC\,3783    &		$10.2^{+3.3 }_{-2.3 }$	&		$6.52	\pm 0.95$  &		$0.104	\pm 0.015 $  \\
NGC\,4051    &		$5.8 ^{+2.6 }_{-1.8 }$	&		$5.10	\pm 0.92$  &		$0.0076	\pm 0.0014 $  \\
NGC\,4151    &		$6.6 ^{+1.1 }_{-0.8 }$	&		$6.2	\pm 3.0$  &		$0.0084	\pm 0.0041 $  \\
PG\,1211+143 &		$93.8^{+25.6}_{-42.1}$	&		$5.06	\pm 0.92$  &		$5.00	\pm 0.91 $  \\
PG\,1226+032 &		$306.8^{+68.5}_{-90.9}$	&		$20.0	\pm 2.6$  &		$86	\pm 11 $  \\
PG\,1229+204 &		$37.8^{+27.6}_{-15.3}$	&		$0.79	\pm 0.23$  &		$0.45	\pm 0.13 $  \\
NGC\,4593    &		$3.7 ^{+0.8 }_{-0.8 }$	&		$7.05	\pm 0.70$  &		$0.0708	\pm 0.0070 $  \\
PG\,1307+085 &		$105.6^{+36.0}_{-46.6}$	&		$1.56	\pm 0.18$  &		$6.59	\pm 0.77 $  \\
IC\,4329A\tablenotemark{b}&$1.5 ^{+2.7 }_{-1.8 }$&		$2.12	\pm 0.73$  &		$0.077	\pm 0.026 $  \\
Mrk\,279     &		$16.7^{+3.9 }_{-3.9 }$	&		$3.88	\pm 0.69$  &		$0.456	\pm 0.082 $  \\
PG\,1411+442 &		$124.3^{+61.0}_{-61.7}$	&		$2.92	\pm 0.32$  &		$3.34	\pm 0.36 $  \\
NGC\,5548    &		$19.7^{+1.5 }_{-1.5 }$	&		$6.0	\pm 1.3$  &		$0.222	\pm 0.047 $  \\
	     &    	$18.6^{+2.1 }_{-2.3 }$	&		$3.4	\pm 1.0$  &		$0.124	\pm 0.037 $  \\
	     &    	$15.9^{+2.9 }_{-2.5 }$	&		$5.52	\pm 0.93$  &		$0.203	\pm 0.034 $  \\
	     &    	$11.0^{+1.9 }_{-2.0 }$	&		$2.8	\pm 1.2$  &		$0.105	\pm 0.043 $  \\
	     &    	$13.0^{+1.6 }_{-1.4 }$	&		$5.19	\pm 0.86$  &		$0.191	\pm 0.032 $  \\
	     &    	$13.4^{+3.8 }_{-4.3 }$	&		$5.9	\pm 1.1$  &		$0.216	\pm 0.040 $  \\
	     &    	$21.7^{+2.6 }_{-2.6 }$	&		$8.2	\pm 1.0$  &		$0.302	\pm 0.037 $  \\
	     &    	$16.4^{+1.2 }_{-1.1 }$	&		$6.7	\pm 1.6$  &		$0.246	\pm 0.060 $  \\
	     &    	$17.5^{+2.0 }_{-1.6 }$	&		$4.25	\pm 0.91$  &		$0.156	\pm 0.033 $  \\
	     &    	$26.5^{+4.3 }_{-2.2 }$	&		$9.6	\pm 1.5$  &		$0.352	\pm 0.054 $  \\
	     &    	$24.8^{+3.2 }_{-3.0 }$	&		$8.0	\pm 1.8$  &		$0.292	\pm 0.067 $  \\
	     &    	$6.5 ^{+5.7 }_{-3.7 }$	&		$3.1	\pm 1.2$  &		$0.114	\pm 0.044 $  \\
	     &    	$14.3^{+5.9 }_{-7.3 }$	&		$3.16	\pm 0.86$  &		$0.116	\pm 0.032 $  \\
	     &    	$6.3 ^{+2.6 }_{-2.3 }$	&		$2.02	\pm 0.36$  &		$0.074	\pm 0.013 $  \\
	     &{\boldmath $18.0^{+0.6 }_{-0.6 }$}&{\boldmath	$3.84	\pm 0.23$} &{\boldmath	$0.205	\pm 0.011 $} \\
PG\,1426+015 &		$95.0^{+29.9}_{-37.1}$	&		$3.48	\pm 0.71$  &		$3.94	\pm 0.81 $  \\
Mrk\,817     &		$19.0^{+3.9 }_{-3.7 }$	&		$4.56	\pm 0.83$  &		$0.56	\pm 0.10 $  \\
	     &    	$15.3^{+3.7 }_{-3.5 }$	&		$3.46	\pm 0.49$  &		$0.423	\pm 0.060 $  \\
	     &    	$33.6^{+6.5 }_{-7.6 }$	&		$3.47	\pm 0.27$  &		$0.424	\pm 0.032 $  \\
	     &{\boldmath $21.8^{+2.4 }_{-3.0 }$}&{\boldmath	$3.54	\pm 0.23$} &{\boldmath	$0.438	\pm 0.028 $} \\
PG\,1613+658 &		$40.1^{+15.0}_{-15.2}$	&		$1.99	\pm 0.43$  &		$5.4	\pm 1.2 $  \\
PG\,1617+175 &		$71.5^{+29.6}_{-33.7}$	&		$1.10	\pm 0.25$  &		$2.30	\pm 0.51 $  \\
PG\,1700+518 &		$251.8^{+45.9}_{-38.8}$	&		$1.85	\pm 0.15$  &		$36.3	\pm 2.9 $  \\
3C\,390.3    &		$23.6^{+6.2 }_{-6.7 }$	&		$0.88	\pm 0.28$  &		$0.44	\pm 0.14 $  \\
Mrk\,509     &		$79.6^{+6.1 }_{-5.4 }$	&		$8.4	\pm 2.0$  &		$1.45	\pm 0.34 $  \\
PG\,2130+099 &    	$22.9^{+4.7 }_{-4.6 }$	&		$4.22	\pm 0.23$  &		$2.51	\pm 0.14 $  \\
NGC\,7469    &		$4.5 ^{+0.7 }_{-0.8 }$	&		$5.14	\pm 0.61$  &		$0.200	\pm 0.023 $  \\

\enddata													      
				    					  					      
\tablecomments{Numbers in boldface are the weighted averages of all						     
	       the measurements for that particular object. Fluxes are						     
	       as observed.  Time lags and luminosities are listed in
	       the rest-frame of the object with weighted averages
	       calculated in log space.}

\tablenotetext{a}{The H$\beta$ lag measurement for this object was
	       deemed unreliable by \citet{peterson04}. In its place,
	       we give the most reliably measured lag, which is for
	       H$\alpha$.}

\tablenotetext{b}{Because of the extremely poor lag measurement (which
               is consistent with zero) and the poor flux calibration
               for this object, we do not include it in the fit to the
               \rl\ relationship.}

\end{deluxetable}

\clearpage

\begin{deluxetable}{lcccc}
\tablecolumns{5}
\tablewidth{0pt}
\tablecaption{H$\beta$ \rl\ Fits}
\tablehead{
\colhead{Note} &
\colhead{N} &
\colhead{K} &
\colhead{$\alpha$} &
\colhead{Scatter\tablenotemark{a}} 
}
\startdata
\multicolumn{5}{c}{FITEXY} \\ \hline \\

All & 59  & $-21.0 \pm 1.8$ & $0.511 \pm 0.041$ & 34.0 \\ 
Avg & 34  & $-22.1 \pm 2.3$ & $0.535 \pm 0.051$ & 40.0 \\
MC  & 34 &$-22.3 \pm 2.2$ & $0.540^{+0.054}_{-0.055}$ & $40.3^{+1.1}_{-0.9}$ \\

\\
\hline
\multicolumn{5}{c}{BCES} \\ 
\hline \\

All & 59 & $-20.4 \pm 1.8$ & $0.499 \pm 0.042$ & \nodata \\
Avg & 34 & $-21.5 \pm 2.1$ & $0.524 \pm 0.046$ & \nodata \\
MC\tablenotemark{b} & 34 & $-21.3^{+2.9}_{-2.8}$ & $0.519^{+0.063}_{-0.066}$ & \nodata \\

\\
\hline
\multicolumn{5}{c}{GaussFit} \\ 
\hline \\

All & 59 & $-21.7 \pm 1.5$ & $0.529 \pm 0.033$ & \nodata \\
Avg & 34 & $-21.8 \pm 1.9$ & $0.531 \pm 0.042$ & \nodata \\
MC  & 34 & $-22.9 \pm 2.2$ & $0.554^{+0.049}_{-0.050}$ & \nodata\\

\enddata

\tablecomments{{\it All:} Each individual measurement is treated
               separately.  {\it Avg:} Multiple measurements for a
               single source are combined into a weighted average.
               {\it MC:} Monte Carlo techniques are used to randomly
               sample the multiple measurements for a single source,
               producing one pair of $R_{\rm BLR}$ and $L$
               measurements per object.  The values and uncertainties
               presented for the fit using this method describe the
               median and 68\% confidence intervals for the
               distributions of slopes and intercepts built up over
               multiple realizations.  As described in the text,
               IC\,4329A was not included in any of the fits
               listed here.}

\tablenotetext{a}{The scatter listed here is the percentage
        of the measurement value of $R_{\rm BLR}$ that is added in
        quadrature to the error value so as to obtain a reduced
        $\chi^2$ of 1.0.}

\tablenotetext{b}{This fit, which properly treats multiple
        measurements of individual objects and accounts for intrinsic
        scatter in the data set, should be taken as our current best
        estimate for the form of the \rl\ relationship.}

\end{deluxetable}

\clearpage

\end{document}